\documentstyle[12pt,epsfig]{article}

\advance\textheight by \topskip
\newcommand{\be}{\begin{equation}}
\newcommand{\ee}{\end{equation}}
\newcommand{\br}{\begin{eqnarray}}
\newcommand{\bea}{\begin{eqnarray}}
\newcommand{\beanon}{\begin{eqnarray*}}
\newcommand{\er}{\end{eqnarray}}
\newcommand{\eea}{\end{eqnarray}}
\newcommand{\eeanon}{\end{eqnarray*}}
\newcommand{\ba}{\begin{array}}
\newcommand{\ea}{\end{array}}
\newcommand{\bi}{\begin{itemize}}
\newcommand{\ei}{\end{itemize}}
\newcommand{\bn}{\begin{enumerate}}
\newcommand{\en}{\end{enumerate}}
\newcommand{\bc}{\begin{center}}
\newcommand{\ec}{\end{center}}

\newcommand{\ar}{\rightarrow}

\newcommand{\Dir}{\kern -6.4pt\Big{/}}%su lettere italiane minuscole
\newcommand{\Dirin}{\kern -10.4pt\Big{/}\kern 4.4pt}
\newcommand{\DDir}{\kern -7.6pt\Big{/}}%su lettere italiane maiuscole
\newcommand{\DGir}{\kern -6.0pt\Big{/}}%su lettere greche
\newcommand{\eeb}{$ e^+e^-$}

\newcommand{\ttb}{$ t\bar t$}
\newcommand{\eett}{$e^+e^-\rightarrow t\bar t $}

\newcommand{\eeHH}{$e^+e^-\rightarrow H^+H^- $}

\newcommand{\eebbww}{$e^+e^-\rightarrow b\bar b W^+W^-$}

\newcommand{\eettbbww}{$e^+e^-\rightarrow t\bar t\rightarrow b\bar b W^+W^-$}

\newcommand{\eebbwh}{$e^+e^-\rightarrow b\bar b W^\pm H^\mp$}
\newcommand{\ttbbwh}{$t\bar t\rightarrow b\bar b W^\pm H^\mp$}
\newcommand{\eettbbwh}{$e^+e^-\rightarrow t\bar t
                              \rightarrow b\bar b W^\pm H^\mp$}

\newcommand{\bbwh}{$b\bar b W^\pm H^\mp$}
\newcommand{\eebbhh}{$e^+e^-\rightarrow b\bar b H^+H^-$}
\newcommand{\ttbbhh}{$t\bar t\rightarrow b\bar b H^+H^-$}
\newcommand{\eettbbhh}{$e^+e^-\rightarrow t\bar t\rightarrow b\bar b H^+H^-$}
\newcommand{\bbhh}{$b\bar b H^+H^-$}
\def\Ord{\buildrel{\scriptscriptstyle <}\over{\scriptscriptstyle\sim}}
\def\OOrd{\buildrel{\scriptscriptstyle >}\over{\scriptscriptstyle\sim}}
\def\sm{\ifmmode{{\cal {SM}}}\else{${\cal {SM}}$}\fi}
\def\mssm{\ifmmode{{\cal {MSSM}}}\else{${\cal {MSSM}}$}\fi}
\def\MH{\ifmmode{{M_{H}}}\else{${M_{H}}$}\fi}
\def\Mh{\ifmmode{{M_{h}}}\else{${M_{h}}$}\fi}
\def\MA{\ifmmode{{M_{A}}}\else{${M_{A}}$}\fi}
\def\MWpm{\ifmmode{{M_{W^\pm}}}\else{${M_{W^\pm}}$}\fi}
\def\MHpm{\ifmmode{{M_{H^\pm}}}\else{${M_{H^\pm}}$}\fi}
\def\Wpm{\ifmmode{{{W^\pm}}}\else{${{W^\pm}}$}\fi}
\def\Hpm{\ifmmode{{{H^\pm}}}\else{${{H^\pm}}$}\fi}
\def\tb{\ifmmode{\tan\beta}\else{$\tan\beta$}\fi}
\def\ctb{\ifmmode{\cot\beta}\else{$\cot\beta$}\fi}
\def\ta{\ifmmode{\tan\alpha}\else{$\tan\alpha$}\fi}
\def\cta{\ifmmode{\cot\alpha}\else{$\cot\alpha$}\fi}
\def\tba{\ifmmode{\tan\beta=1.5}\else{$\tan\beta=1.5$}\fi}
\def\tbb{\ifmmode{\tan\beta=30.}\else{$\tan\beta=30.$}\fi}
\def\cab{\ifmmode{c_{\alpha\beta}}\else{$c_{\alpha\beta}$}\fi}
\def\sab{\ifmmode{s_{\alpha\beta}}\else{$s_{\alpha\beta}$}\fi}
\def\cba{\ifmmode{c_{\beta\alpha}}\else{$c_{\beta\alpha}$}\fi}
\def\sba{\ifmmode{s_{\beta\alpha}}\else{$s_{\beta\alpha}$}\fi}
\def\ca{\ifmmode{c_{\alpha}}\else{$c_{\alpha}$}\fi}
\def\sa{\ifmmode{s_{\alpha}}\else{$s_{\alpha}$}\fi}
\def\cb{\ifmmode{c_{\beta}}\else{$c_{\beta}$}\fi}
\def\sb{\ifmmode{s_{\beta}}\else{$s_{\beta}$}\fi}
\def\pl #1 #2 #3 {{\it Phys.~Lett.} {\bf#1} (#2) #3}
\def\np #1 #2 #3 {{\it Nucl.~Phys.} {\bf#1} (#2) #3}
\def\zp #1 #2 #3 {{\it Z.~Phys.} {\bf#1} (#2) #3}
\def\pr #1 #2 #3 {{\it Phys.~Rev.} {\bf#1} (#2) #3}
\def\prep #1 #2 #3 {{\it Phys.~Rep.} {\bf#1} (#2) #3}
\def\prl #1 #2 #3 {{\it Phys.~Rev.~Lett.} {\bf#1} (#2) #3}
\def\mpl #1 #2 #3 {{\it Mod.~Phys.~Lett.} {\bf#1} (#2) #3}
\def\rmp #1 #2 #3 {{\it Rev. Mod. Phys.} {\bf#1} (#2) #3}
\def\xx #1 #2 #3 {{\bf#1}, (#2) #3}
\def\preprint{{\it preprint}}
\begin{document}
\tolerance=100000
\thispagestyle{empty}
\setcounter{page}{0}

\begin{flushright}
{\large Cavendish--HEP--97/01}\\
{\rm January 1997\hspace*{.5 truecm}}\\ 
\end{flushright}

\vspace*{\fill}

\begin{center}
{\large \bf 
The processes $e^+e^-\rightarrow b\bar bW^\pm H^\mp$ 
and $e^+e^-\rightarrow b\bar bH^+H^-$ \\
at the Next Linear Collider\\
in the Minimal Supersymmetric Standard Model}\\[2.cm]
{\large Stefano 
Moretti$^1$ and Kosuke Odagiri\footnote{E-mails: 
moretti,odagiri@hep.phy.cam.ac.uk}}\\[.3 cm]
{\it Cavendish Laboratory,
University of Cambridge,}\\
{\it Madingley Road,
Cambridge, CB3 0HE, United Kingdom.}\\[1cm]
\end{center}

\vspace*{\fill}

\begin{abstract}
{\normalsize
\noindent
The complete matrix elements for $e^+e^-\rightarrow b\bar bW^\pm H^\mp$ 
and $e^+e^-\rightarrow b\bar bH^+H^-$ are
computed at tree-level within the Minimal Supersymmetric
Standard Model. 
Rates of interest to phenomenological analyses at the Next Linear
Collider are given. In particular, we analyse: (i)
$t\bar t$ pair production followed by the top
decays into $b\bar bW^\pm H^\mp$ and $b\bar b H^+H^-$
final states; (ii) $H^+H^-$ signals in which 
$H^\pm\ar h W^{\pm} \ar b\bar b W^{\pm}$ and $H^\pm\ar tb\ar 
b\bar b W^\pm$. Top and Higgs finite width effects are included
as well as all those due to the irreducible backgrounds.}
\end{abstract}

\vspace*{\fill}
\newpage

\section*{1. Introduction} 

In the present paper we intend to conclude a long-term project
of computing effects due to the finite width of the unstable particles
as well as to the non-resonant diagrams entering in electroweak (EW)
processes of the type
\be\label{process}
e^+e^-\ar b\bar b X^+ X^-,
\ee  
where $X^\pm=W^\pm,H^\pm$, within the Standard 
Model (\sm) and the
Minimal Supersymmetric Standard Model (\mssm) at Next Linear Collider
(NLC) energies \cite{ee500}. The case $X^+X^-\equiv W^+W^-$ 
within the \sm\ was considered
in Refs.~\cite{tt,ZH} (see also Ref.~\cite{noi5}) whereas the equivalent 
process in the \mssm\ was discussed in Ref.~\cite{io}\footnote{For the case of 
the irreducible background from QCD 
in $W^\pm+\mbox{4jet}$ events, see Ref.~\cite{W4j}.}.

By the time that an NLC machine will be operating, 
$m_t$ will already be known precisely and either $M_\phi$ 
will have been measured or a large part of the \mssm\ parameter space 
$(\MA,\tb)$ defining the values of $M_{\rm{Higgs}}$ 
will have already been explored\footnote{Here and in the following, $\phi$
indicates the Higgs boson of the \sm\ whereas the subscript
`Higgs' refers 
collectively to the ${H,h,A,H^\pm}$ scalars of the \mssm.}, thanks
to the experiments at the Tevatron \cite{Teva}, 
LEP2 \cite{lep2w} and LHC \cite{CMS,ATLAS}. 
(For an illustration of the discovery potential of such machines, we
refer the reader to \cite{Teva,lep2w,CMS,ATLAS}.) Anyhow,
whichever the status of the measurements of $m_t$, 
\MH, \Mh, \MA\ and/or \MHpm\ will be, we would like to stress that
detailed studies of both top quark and Higgs
boson properties (other than the mass, e.g., the width, the branching 
ratios, etc.)
will more easily be carried out at an \eeb\ linear accelerator, operating
in the energy range $2m_t \Ord \sqrt s\Ord 1$ TeV. At such a machine, 
the absence of the huge QCD background typical of 
hadron colliders combined with the large luminosity and `effective' 
centre-of-mass (CM)
energy available allows one to cover the whole of the Higgs mass range
and to exploit all possible Higgs and top decay channels in order
to perform very high precision measurements of their parameters.

Top and Higgs signatures naturally enter in processes of the type 
(\ref{process}). In fact, in the \sm, the process \eebbww\ 
is a signature of both top quark production
in $t\bar t$ pairs and Higgs boson bremsstrahlung 
in the $Z\phi$ channel. On the one hand, top pairs produced
via the \eett\ decay 
through $t\bar t\ar b\bar bW^+W^-$ whereas, on the
other hand, the channel $Z\phi\ar b\bar bW^+W^-$ will be one of 
the cleanest
ways of detecting a heavy Higgs, thanks to the high performances expected 
from the
vertex detectors in triggering $b$-quarks \cite{ideal}\footnote{The 
branching ratio (BR) of the channel
$Z\ar b\bar b$ is 
five times larger than that into $\mu^+\mu^-$
or $e^+e^-$ and it is free from backgrounds due to $W^\pm$ decays.}.
Within the \mssm, the reaction \eebbww\ allows one to study 
not only the Supersymmetric counterparts of the two mentioned \sm\ 
processes\footnote{In case of Higgs production, the heaviest of the \mssm\
neutral Higgses (i.e., $H$) replacing the \sm\ Higgs scalar (i.e., $\phi$).},
but also $AH$ production followed by $A\ar b\bar b$ and $H\ar W^+W^-$
as well as $hW^+W^-$ in which $h\ar b\bar b$ \cite{io}.

It is the purpose of this note to consider the last two possibilities left
within the \mssm, that is 
\be\label{wh}
e^+e^-\ar b\bar b W^\pm H^\mp,
\ee
\be\label{hh}
e^+e^-\ar b\bar b H^+ H^-.
\ee
As it was done in the previous publications \cite{tt}--\cite{io}, 
our aim is twofold.
\begin{itemize}
\item Firstly, to establish the importance of top finite width 
effects and of those due to the non-resonant
background in case of \ttbbwh\ and \ttbbhh\ decays.
\item Secondly, to study Higgs 
signatures, such as $e^+e^-\ar H^+H^-$ production followed by the decay
of one of the charged Higgses into:  
\end{itemize}
\begin{enumerate}
\item $h^{*}W^{\pm *}\ar b\bar b W^{\pm *}$ and 
\item $t^{*}b\ar b\bar b W^\pm$ (if $m_t\Ord M_{H^\pm}-m_b$),

\noindent
where the asterisk indicates possible off-shell channels.
Note that these two signatures are produced via process
(\ref{wh}).
We further notice that the top-antitop decay channels
\item \ttbbwh\ and \ttbbhh\ 

\noindent
represent themselves a source of $H^\pm$ signals.
\end{enumerate}

Although all the topologies yielding \bbwh\ and \bbhh\ final states 
produce Higgs bosons (both charged an neutral), we will consider the
corresponding diagrams as (irreducible) `background' to
the two `signals' \eett\ and \eeHH, as the former represent Higgs production
mechanisms that are largely suppressed with respect to the latter
and also to others exploited in neutral Higgs boson searches (for a review
of these see, e.g., Ref.~\cite{Higgsreview})\footnote{We notice 
that such diagrams are nonetheless interesting on their own, e.g., as a way 
of testing the gauge structure of Higgs sector of the \mssm, 
as they involve many of the fundamental  tree-level vertices of the theory.
This is however beyond the scope of this study.}. 
Furthermore, it is worth
reminding 
the reader that, at least over the areas of the \mssm\ parameter space
where the $t\ar bH^+$ and $H^+\ar hW^+\ar b\bar b W^+$ decay modes
are both important (see later on), the processes \eett\ and \eeHH\ 
cannot be unambiguously separated and studied independently.
Therefore, in such $(\MA,\tb)$ regions, 
any of the two reactions constitutes an irreducible background to the other.
The different topologies contributing at leading order to processes
(\ref{wh})--(\ref{hh}) are given in Figs.~1a and 1b, respectively.
In the first case they lead to 51 different Feynman diagrams whereas
for the second case the number is 60.

For future reference, we also present in Fig.~1c the \mssm\ BRs
of top quarks and charged Higgs bosons in the $M_{H^\pm}$ range and for the 
values of $\tan\beta$ that will be used in our analysis. 
Note that such rates dictate the decay phenomenology of the 
top (diagrams 2 in Fig.~1a and 3 in Fig.~1b) and Higgs 
(diagrams 9 and 12 in Fig.~1a) signals that will be studied later on.
As in our forthcoming discussions we will interchange the r\^ole of $M_{H^\pm}$
and $M_A$, we have plotted both these quantities as mass scales in Fig.~1c. 

The plan of the paper is as follows. In the next Section we describe
the calculations we have performed. In Section 3 we present
our results. Our conclusions are given in Section 4.

\section*{2. Matrix Elements}

To compute the Matrix Elements (MEs) 
of processes (\ref{wh})--(\ref{hh}) we have
used helicity amplitude techniques. A first code has been produced by using
the method of Ref.~\cite{HZ}. This has then been checked against
a second one built up by resorting to the subroutines contained in the package
HELAS \cite{HELAS}. In particular, 
in order to implement the diagrams of Figs.~1a--b,  we
have made use of the subroutines involving scalars with the
vertices and propagators appropriately arranged to reproduce the
interactions and wave-functions of the five Higgses in the unitary gauge of the
\mssm.

The two {\tt FORTRAN} codes have been programmed
by the two authors independently and they agree at amplitude squared level
within 12 significative digits in {\tt REAL*8} precision. Moreover, all the
amplitudes produced with the first method have been
tested for gauge invariance by computing the two MEs in other two gauges
as well (i.e., the Feynman and the Landau ones). In order to deal
with the complicated resonant structure of processes (\ref{wh})--(\ref{hh}),
we have adopted the technique of splitting the two MEs 
in a sum of non-gauge-invariant pieces, each of these implementing a 
different resonant structure, and of integrating them separately with
the appropriate mapping of the phase space variables. Such a procedure
has already been documented in detail for the case of the \eebbww\ process 
in the \sm\ in Ref.~\cite{ZH}, so we only sketch out here the method.

Firstly, one isolates the diagrams with similar resonant structure by
grouping these together in `subamplitudes'. For example, one can recognise
in process (\ref{wh}) the resonances: $t\bar t\ar
(bW^+)(\bar b H^-)$, $t\ar bW^+$, $\bar t\ar \bar b H^-$, $H,h,A,Z\ar b\bar b$,
$H^+H^-\ar (t\bar b) H^-\ar
(b\bar b W^+)H^-$ and $H^+H^-\ar (hW^+) H^-\ar (b\bar b W^+)H^-$. 
In case of reaction (\ref{hh}) one has the following Breit-Wigner peaks: 
$t\bar t\ar(bH^+)(\bar b H^-)$, $t\ar bH^+$, $\bar t\ar \bar b H^-$ and 
$H,h,A,Z\ar b\bar b$.
Secondly, one defines the mentioned non-gauge-invariant components of
the amplitudes squared, by appropriately combining the subamplitudes.
For example, in both processes (\ref{wh})--(\ref{hh}) we have simply taken the 
square of the various resonant subamplitudes (yielding 9 and 7 terms, 
respectively) whereas we have combined their interferences altogether
in the last term, apart from those between the two 
single-top and the only double-top channels, which were added to the
$t$- and $\bar t$-diagrams in both reactions (\ref{wh})--(\ref{hh}).
Thirdly, one maps the phase spaces around the resonances. For example,
for our analysis this was made by performing the change of variable
\be\label{change}
Q^2-M^2=M\Gamma\tan\theta, \longrightarrow
{{d}}Q^2=\frac{(Q^2-M^2)^2+M^2\Gamma^2}{M\Gamma}{{d}}
\theta
\ee
(where $Q$, $M$, $\Gamma$ stand for the virtuality, the mass and the width
of the resonance), which gives integrands smoothly dependent on $\theta$.
Fourthly, the various amplitude squared terms are integrated separately 
(we have used VEGAS \cite{VEGAS} to this purpose)
and added up in the end (to recover gauge invariance).
We however notice that, if taken
separately, these terms provide  a useful way of
looking inside the processes and distinguishing between the different
fundamental interactions.

Concerning the values adopted for 
the \sm\ and \mssm\ parameters and the
relations implemented for the \mssm\ Higgs masses and couplings 
we refer to Refs.~\cite{tt}--\cite{io}.
Furthermore, 
as in the previous publications \cite{tt}--\cite{io}, 
we have chosen here, as representative for
$\tan\beta$, the two extreme values
1.5 and 30. $M_{A}$ varies between 
50 and 220 GeV, as higher values would be uninteresting in the
present context (the corresponding rates for both the $t\ar bH^\pm$ 
decays and the $H^+H^-$ pair production become heavily suppressed). Following
the latest measurements at FNAL \cite{topmass}, 
the top mass has been set equal to 174 GeV. Therefore, we use as indicative of
the top-antitop threshold energy the value $\sqrt s=350$ GeV, along with
the final stage energy of 500 GeV.

Our calculations have been carried out at leading order only. 
We have not included beamsstrahlung and bremsstrahlung effects either. 
We are however confident that such approximations will not spoil our 
conclusions. Indeed, it has been proved that this is 
true for the process \eebbww\ \cite{tt,io} and we expect the same
in the present context.

A final consideration is now in order. Following the approach of 
Refs.~\cite{tt}--\cite{io}, we have generally kept the final state bosons 
$W^\pm$ and $H^\pm$ on-shell in our computations. However, we 
made an exception in one case: when computing the squared of graph 12
of Fig.~1a. This represents the Higgs signal
$e^+e^-\ar H^+H^-$ in which one of the two scalars decays into
$h W^\pm $ pairs, followed by $h\ar b\bar b$. That is, when only
the light neutral Higgs is allowed to be off-shell. However, this channel
is relevant in the case of off-shell $W^{\pm }$ bosons as well. 
Therefore, in order to keep this into account, 
when calculating the contribution of the mentioned diagram
to the total cross section of process (\ref{wh}) we have 
attached a fermion-antifermion decay current to the $W^\pm$ boson
and computed the amplitude squared of a 2$\ar$5 process.
Indeed, proceeding this way, the rates that we calculated from the
integration of this diagram over the appropriate phase space
reproduce rather accurately those obtained by multiplying the cross section
for \eeHH\ \cite{Higgsreview} 
times the off-shell BR$[H^{\pm}\ar hW^{\pm } \ar 
b\bar b\mbox{jj}]$ \cite{ioejames}, where j represents
a jet (see next Section).

\section*{3. Results}

In carrying out our analysis we closely follow 
Refs.~\cite{tt,ZH,io} and the studies reported in the Sections 
`Higgs Particles' and `Top Quark Physics' of Ref.~\cite{ee500}. 
Furthermore, throughout this paper
we assume that the mass scale of the Supersymmetric partners
of ordinary matter is beyond the energy reach
of the NLC, so that they cannot be produced at such a collider.
In particular, we neglect considering the Supersymmetric decay
$t\ar \tilde{t}\tilde{\chi}_1^0$, where $\tilde{t}$ refers to the
stop and $\tilde{\chi}_1^0$ is the lightest
neutralino, as well as other possible \mssm\ modes\footnote{For a
recent review of the latter, see Ref.~\cite{Guasch}.} \cite{Bagliesi}.

\subsection*{3.1 Event selection}

In our selection strategy we assume a six-particle
semi-leptonic(hadronic) tagging, for both reactions (\ref{wh})--(\ref{hh}):
that is, $b$'s are assumed to hadronise whereas one of the charged bosons 
decays into jet-jet pairs and the other into $\tau\nu_\tau$.
In this way, the selected final state is always made up by   
a high energy and isolated $\tau$ accompanied by a multi-hadronic system 
and appreciable missing four-momentum. 
\vfill\newpage
\noindent\underline{\boldmath${e^+e^-\ar b\bar b W^\pm H^\mp}$}
\vskip0.25cm
\noindent
This means that in events of the type (\ref{wh}) one can in principle 
pick up a tau-lepton either from a $W^\pm$ or a $H^\pm$. However,
it should be noticed that the four-momentum of the tagged boson 
is eventually reconstructed by summing that of the
decay lepton to the missing one (the latter being computed using the 
kinematic constraints \cite{Bagliesi})\footnote{This is exactly true 
only when the energy of the colliding 
electron/positron beams is known
(i.e., no Linac energy spread,  bremsstrahlung and beamsstrahlung), when
the detector is ideal (i.e., infinite resolution and complete
geometrical coverage) and when no other neutrino from cascade decays
is present in the interaction. In a more realistic case, corrections
have to be made to account for all the above effects, which in fact do not
spoil the effectiveness of the method \cite{Bagliesi}.}. Therefore, 
one can always
ask that the invariant mass of the lepton-neutrino system is larger than
$\sqrt{M_{W^\pm}^2+M_A^2}\approx 100$ GeV to 
remove $W^\pm\ar \tau\nu$
decays, $\MA\approx60$ GeV being the lower experimental limit on the mass
of the pseudoscalar neutral Higgs of the \mssm\ \cite{limA}. 
This way, we explicitly 
require only hadronic $W^\pm$ decays in process (\ref{wh}). This has a 
clear advantage: in fact, it allows one
to easily reconstruct resonant invariant mass
spectra (i.e., $M_{bW^\pm\ar b\rm{jj}}$ for the top and 
$M_{b\bar bW^\pm\ar b\bar b\rm{jj}}$
for the charged Higgs).
In addition, the hadronic channels
represent the largest fraction of the total $W^\pm$ decay rate
(around 70\%). 
We also believe the selection
of the channel $H^\pm\ar\tau\nu_\tau$ in reaction (\ref{wh})
not restrictive in the present
context, for two reasons.
On the one hand, the tau-neutrino channel is always sizeable up to 
$M_{H^\pm}\approx 170-180$ GeV (i.e., the $tb$ decay threshold) and
for all $\tb$'s.  
\cite{ioejames}. 
Therefore, when tagging semi-leptonically(hadronically) over this 
$M_{H^\pm}$ range
one naturally retains a large part of the original top (diagram 2 in Fig.~1a)
and Higgs (diagram 12 in Fig.~1a) events. Indeed, it should be
noticed that only charged Higgses with mass up to 161 GeV or so 
(corresponding to $\MA\approx140$ GeV) can
be produced with detectable rate at $\sqrt s=350$ GeV via \eeHH. 
On the other hand, when $M_{H^\pm}\OOrd 180$ GeV (thus when
$\sqrt s=500$ GeV), top quarks mainly decay via
the $b\bar W^\pm$ channel so that \eett\ events are in practise accounted 
for by the \eebbww\ channel alone \cite{tt,io}. In addition, we remark that
double $t$- and/or $h$-decays $H^+H^-\ar 
b\bar bt\bar t, tb hW^\pm, hh W^+W^- \ar b\bar bb\bar b W^+W^-$
of charged Higgses produced via \eeHH\ would yield a final signature
with numerous hadronic tracks (e.g., up to 8 jets 
can be produced, if $W^+W^-\ar \mbox{jjjj}$ !) with the consequent
problems due to the combinatorics, so that to tag one
of the Higgses via the lepton-neutrino channel might be a better 
approach\footnote{Alternatively, such multi-hadronic channels would need 
a proper simulation (including hadronisation, detector effects, etc.)
which is clearly beyond the intention of this study.}.
For example, for large $\tb$ values, the $H^\pm\ar\tau\nu_\tau$ channel 
has a BR that is always larger than $\approx0.35$  \cite{ioejames}
for the values of $H^\pm$ masses that can be produced at
$\sqrt s=500$ GeV (say, up to $M_{H^\pm}\approx235$ GeV, that is, for
$\MA\Ord 220$ GeV). In contrast, if $\tb$ is small, the heavy mass region
$M_{H^\pm}\OOrd m_t$ is clearly sacrificed by requiring one of the
$H^\pm$'s to decay into $\tau\nu$ pairs \cite{ioejames}, so that in such a
case the mentioned multi-hadronic final states will necessarily need to be 
exploited in experimental analyses. 
\vskip0.5cm
\noindent\underline{\boldmath${e^+e^-\ar b\bar b H^+H^-}$}
\vskip0.25cm
\noindent
In the case of reaction (\ref{hh}) one is only interested in top decays 
(diagram 3 of Fig.~1b),
since $H^\pm$ events cannot be produced at large rate by any of the other 
diagrams. Therefore, in this case, the relevant charged Higgs masses 
are those for which $M_{H^\pm}< m_t-m_b$ and  strictly below the
top-bottom threshold the $\tau\nu_\tau$ 
channel is the dominant one at all $\tb$'s  \cite{ioejames}. 
In these conditions
then, a purely leptonic decay channel might seem more appropriate: i.e.,
$t\bar t\ar b\bar b H^+H^-\ar b\bar b\tau^+\tau^-\nu_\tau\bar\nu_\tau$.
However, we do not consider here this possibility. Firstly, because
it would prevent one from reconstructing invariant mass top/Higgs spectra and,
secondly, because over the experimentally allowed $\tb$ range 
(say, between 1 and 40) the
$t\ar bH^\pm$ decays are always suppressed compared to the
$t\ar bW^\pm$ ones \cite{DKZ}, so that the $t\bar t\ar b\bar b W^\pm H^\mp$
channel represents a better option anyway to detect 
charged Higgs scalars \cite{Higgsreview}.
\vskip0.5cm
\noindent
Our results are presented in Tabs.~I--II and Figs.~2--8\footnote{Please
note that, when reporting the rates corresponding to process (\ref{wh})
in the forthcoming Figures and Tables,
we have considered only one of the two possible charge combinations of the
$W^\pm$ and $H^\mp$ bosons. The cross sections for 
$e^+e^-\ar b\bar b W^+H^-$ and $e^+e^-\ar b\bar b H^+W^-$ are in fact
identical because of charge conjugation. Therefore, the total production rates 
correspond to twice the values given in Tabs.~I--II and Figs.~2--8. 
We have decided to do so in order to be able to plot in the same figures
the spectra of both processes (\ref{wh})--(\ref{hh}), the rates of
the latter being generally smaller than those of the former.}.

\subsection*{3.2 Top signals}

As the techniques to measure the top parameters differ greatly depending
on whether the collider is running around the $2m_t$ threshold or far above 
that, we will discuss separately the main features of processes
(\ref{wh})--(\ref{hh}) at the two CM energies considered.
\vskip0.5cm
\noindent\underline{{\boldmath${\sqrt s=350}$}~{\bf{GeV}}}
\vskip0.25cm
\noindent
Since one of the means of measuring $m_t$ at an NLC is a threshold scan
\cite{Bagliesi},  Figs.~2a and 2b
show the total cross sections in an interval around the top-antitop 
threshold at $2m_t$ for the case of \bbwh\ and \bbhh\ final states, 
respectively,
for a selection of pseudoscalar Higgs masses (for larger values of $\MA$
top decays into charged Higgs are well below detection level): in the
case of the \eett\ diagram alone as well as in presence of all graphs.
Whereas in the case of two charged Higgses (see Fig.2b) the irreducible
background is almost negligible (only a ten percent effect at the most
can be noted very below threshold, at small $\MA$'s), when only one of 
the top quarks decays
into $bH^\pm$ pairs its effect can be drastically  visible (see Fig.~2a).
This is always true at small $\tb$, whereas for large $\tb$'s this is the 
case only at smaller 
values of $\MA$ (for $\MA=140$ GeV and $\tbb$ the $t\bar t$ and the
complete \bbwh\ rates virtually coincide).
To appreciate quantitatively the various effects entering in the curves
of Fig.~2a--b we have displayed in three different columns
(see Tabs.~I and II) the cross sections due
to: (i) the \eett\ diagram in Narrow Width Approximation (NWA,
$\Gamma\ar0$)\footnote{See 
Refs.~\cite{tt,io} for details.}; (ii) the \eett\ diagram with a finite
width 
($\Gamma=\Gamma_t$); (iii) all the diagrams (again with $\Gamma=\Gamma_t$)
for both processes (\ref{wh})--(\ref{hh}) and the two 
$\tb$ values. This
allows one to isolate the effect of the finite width $\Gamma_t$
and of the irreducible background. We also report 
rates for three values of $M_{H^\pm}$ so that one can notice the heavy
suppression of the total cross sections, which sets up 
as the charged Higgs mass increases and which is due to the decreasing
branching ratio of the top into $bH^\pm$ pairs.
 
Thanks to our procedure of computing the `total cross sections' (i.e., by
summing different `sub-cross sections' with a specific resonant
structure) we have been able to identify which of the diagrams in 
Figs.~1a--b are responsible for the larger background contributions
entering in Figs.~2a. 

For example, for $\tba$, at $\MA=60$ GeV, $h(A)\ar bb$
resonant graphs (not including graph 12 in Fig.~1a) yield a contribution
of $\approx 0.16(0.41)$ fb, at $\sqrt s=340$ GeV. Of the
two Higgs signals only $H^\pm\ar hW^{\pm }\ar b\bar bW^{\pm }$   is
relevant, producing a rate of $\approx8.68$ fb, whereas $H^\pm\ar tb\ar
b\bar bW^\pm$ is around 0.16 fb only ($M_{H^\pm}\approx100$ GeV being a value
well below the threshold at $m_t+m_b$). The single top resonant diagrams
via $t\ar bW^\pm$ (the square of diagram 6 plus 
its interference with graphs 2 and 9 in Fig.~1a) and via 
$t\ar bH^\mp$ (the square of the sum of 
diagrams 1, 4 and 5 plus their interferences with graph 2 in Fig.~1a)
produce the rates 0.10 and 0.46 fb, respectively.
At the same point of CM energy the \ttb\ contribution is $\approx2.47$ fb.
As $\sqrt s$ increases, the top-antitop signal clearly gets larger, eventually
being the dominant part of the total rate: e.g., $\approx72$ fb at 360 GeV. 
The Higgs resonances $h$ and $A$ yield 0.19 and 0.52 fb at $\sqrt s=360$ GeV, 
respectively, whereas the single-top ones produce
0.18 and 0.65 fb, in correspondence of $t\ar b W^\pm$ and
$t\ar b H^\mp$ diagrams.
The trend remains rather similar for $\MA=100$ GeV (and $\tba$), whereas
when $\MA=140$ GeV (that is $M_{H^\pm}\approx161$ GeV) phase space suppression
effects become dominant. On the one hand, the off-shell $t\bar t$ production
is well below detection level up to $\sqrt s\approx346$ GeV. On the other
hand, the $h(A)\ar b\bar b$ resonant graphs are much 
smaller (their BRs remain constant but
the sum $M_{h(A)}+M_{H^\pm}+M_{W^\pm}$ approaches(exceeds) the CM energy).
Therefore, the bulk of the total cross section is due
to the $H^\pm$ signals, $H^\pm\ar hW^{\pm }\ar b\bar b W^{\pm }$ being 
around 2.19 fb and $H^\pm\ar tb\ar b\bar b W^\pm$ approximately 3.45 fb
(as $M_{H^\pm}\ar m_t+m_b$), again at $\sqrt s=340$ GeV.
The phase space effect is very drastic on top decays, so that the
difference between the \eett\ curve and that produced by all the diagrams is
largest in this case (i.e., $\MA=140$ GeV). 
In fact, an increasing value of $M_{H^\pm}$ means a suppression of the
BR$[t\ar bH^\pm]$ and vice versa an enhancement of that for
$H^\pm\ar tb\ar b\bar bW^\pm$ (also note that for $M_{H^\pm}\approx161$
GeV one gets closer to a local maximum of the $H^\pm\ar hW^{\pm }\ar b\bar b
W^{\pm }$ decay fraction \cite{ioejames}). 

For $\tbb$ the two $H^+H^-$ signals are generally negligible, because of the
suppression coming from their BRs (see Ref.~\cite{ioejames}), the
dominant background effects being due to the $h\ar b\bar b$ and $A\ar b\bar b$
resonant graphs. At $\sqrt s=340$ GeV and for $\MA=60$ GeV, one gets
that $H^\pm\ar hW^{\pm }\ar b\bar bW^{\pm }$ and $H^\pm\ar 
tb\ar b\bar bW^\pm$ are
at the level of ${\cal O}(10^{-2})$ and ${\cal O}(10^{-4})$ fb, respectively, 
whereas the $h(A)\ar b\bar b$ resonant graphs yield a contribution
of 0.39(0.40) fb. However,
as $\MA$  increases the $h,A\ar b\bar b$ resonances get
smaller (they both are
around 0.015 fb at $\MA=100$ GeV and both completely negligible
at $\MA=140$ GeV, again for $\sqrt s=340$ GeV). In fact, although
the BRs for $h$ and $A$ bosons into $b\bar b$ pairs remain roughly 
constant for $\MA\Ord140$ GeV also for \tbb, it should be noticed that 
$\Mh$ grows with $\MA$ more at large than at small \tb's. For example, at
$\MA=100(140)$ GeV and \tbb, one finds 
$M_{h}+M_{W^\pm}+M_{H^\pm}\approx 308(336)$ GeV. For such reasons then, the
differences between \eettbbww\ and \eebbww\ rates are small for
$\MA\OOrd100$ GeV and large $\tb$'s.

A second technique of measuring $m_t$ at threshold
is the so-called end-point method in the spectrum of
the boson produced in the top decay chain \cite{Bagliesi}.
Whereas for events of the type \eebbww\ this is necessarily a $W^\pm$ boson
\cite{tt,io},
according to our detection strategy such a particle is in 
events of the type (\ref{wh})--(\ref{hh}) a $H^\pm$ scalar,
tagged in the $\tau\nu_\tau$ channel. 
In general, if one assumes an NWA for
the top in \eett\ (see eq.~(1) in Ref.~\cite{tt}), then the minimum and 
maximum values of $p_H$ (and $p_W$) are fixed by the decay kinematics (see,
e.g., \cite{By}). This is clearly 
spoiled by a finite value of $\Gamma_t$ (which introduces a Breit-Wigner
dependence in the top propagator), whose effect consists in
a smearing of the edges
of the momentum distribution. The inclusion of all the other non-\ttb\
diagrams also mimic a similar effect. The $p_{H}$ spectrum in presence
of both finite top width and all graphs of Figs.~1a--b is given in 
the two upper
windows of Figs.~3a--b (e.g., for $\MA=60$ and 100 GeV). For reference, we 
also give the same distribution in the
case of the $W^\pm$ boson (from \eebbwh\ events), spectrum which 
could well
be plotted by means of the momentum of the jet-jet pair for which 
$M_{\rm{jj}}\approx M_{W^\pm}$.
From their comparison with the curves in NWA (lower windows) one can
easily deduce the difficulty in recognising any end-point in the 
distributions when all effects are properly included, 
those more visible (the long tails at large momentum for \eebbwh) being clearly
due to the irreducible background. 
It is also interesting to notice the single bin filled on the right hand side 
of the two upper plots in Fig.~3a, 
for the case of the $p_H$ spectrum in reaction (\ref{wh}). This is
due to the \eeHH\ diagrams (graphs 9 and 12 in Fig.~1a), for which
the momentum of the $H^\pm$'s is fixed via the relation 
$p_H=\sqrt {s/4 -M_{H^\pm}^2}$. The rate of this bin is regulated
by the interplay between the cross section for \eeHH\ and the charged
Higgs BRs into $hW^{\pm}$ and $tb$ pairs, so that
the bin practically disappears for \tbb, for which the 
latter are both heavily suppressed (when $\MA\Ord100$ GeV).
\vskip0.5cm
\noindent\underline{{\boldmath${\sqrt s=500}$}~{\bf{GeV}}}
\vskip0.25cm
\noindent
Far above the $2m_t$ threshold (in our case, $\sqrt s=500$ GeV)
one can determine the top mass from the two three-jet invariant mass 
distributions that can be reconstructed
from the $b\bar b\rm{jj}\tau\nu_\tau$ 
final state \cite{Bagliesi}. In fact, in \bbwh\ and \bbhh\ events
in which one final state boson decays hadronically (let us say the positive
charged one $B^+$, where $B^\pm=W^\pm,H^\pm$)
and the other leptonically, there are two possible three-jet
combinations among the $\rm{jj}$ pair reconstructing the boson mass
(i.e., those for which $M_{\rm{jj}}\approx M_{W(H)^\pm}$)
and the two $b$'s: the `right' one ($B^+b$), peaking at $m_t$, and the
`wrong' one ($B^+\bar b)$, producing a rather broad and flat 
spectrum  (we 
assume that the charge of the $b$-jets is not recognised). They are 
both plotted in Figs.~4a and b, for both reactions
(\ref{wh})--(\ref{hh}). For comparison, in case of \bbwh\ events we also 
give the similar distribution which can be obtained by combining the 
tagged $H^\pm$ with the two bottom quarks. Note
that by `$b$-quarks' we mean here the two jets that do not peak
at $M_{W(H)^\pm}$\footnote{The tagging procedure can be
simplified if $\mu$-vertex tagging techniques are exploited on
the two $b$-quarks, so that
the requirement $M_{\rm{jj}}\approx M_{W(H)^\pm}$ needs not to be
applied.}. Note the logarithmic scale of the figures, as well as
the rather thin binning (2 GeV) of the distributions. In other terms,
although in this case
the `wrong' distributions are well below the 
`right' ones, it must be remembered that for resolutions worse than
that considered here
the peaks would be cut down, whereas the flat spectra would remain 
the same. Therefore, the possibility of clearly identifying 
the top resonance drastically depends on the angle and energy resolution
of each single jet. 

For reference, the cross sections at $\sqrt s=500$
GeV for process (\ref{wh}) are (no charge conjugation included): 
105(114) fb for $\tba$ and 61(64) fb for $\tbb$ when $\MA=60$ GeV;
  61(76) fb for $\tba$ and  35(37) fb for $\tbb$ when $\MA=100$ GeV.
At the same energy, in case of reaction (\ref{hh}) one obtains:
 24(24)  fb for $\tba$ and 8.5(8.6) fb for $\tbb$ when $\MA=60$ GeV;
6.4(6.5) fb for $\tba$ and 2.3(2.4) fb for $\tbb$ when $\MA=100$ GeV.
The numbers in brackets refer to the rates obtained when all diagrams
are calculated whereas the others to the case of top-antitop production
and decay only (in both cases $\Gamma=\Gamma_t$). From these numbers, one 
realises that in case of \eebbhh\ events the effects 
of the irreducible 
background are negligible, whereas for the process \eebbwh\  they can 
amount to several percent (also at $\sqrt s=500$ GeV).

\subsection*{3.3 Higgs signals}

Let us now turn to the case of Higgs resonances. We have chosen only a
few representative cases, where Higgs peaks can be visible in the 
decay spectra and show peculiar features. First of all, it is instructive to 
plot the invariant mass of the $b\bar b$ system, when all diagrams are
computed. This is done in Fig.~5a for the 
combination $\MA=60$ GeV and \tba, at $\sqrt s=350$ GeV, for both
processes (\ref{wh})--(\ref{hh}),
and in Fig.~5b for the 
case $\MA=140$ GeV and \tba, at $\sqrt s=500$ GeV, for reaction (\ref{wh})
only. If the resolution in invariant mass of the $b\bar b$ system is 5 GeV,
at lower energy
only the spectrum from process (\ref{wh}) shows a resonance (see Fig.~5a,
continuous histogram in the main plot).
However, this peak actually hides a superposition of two
$M_{b\bar b}$ resonances, due to decaying $h$ and $A$ bosons (in order
of increasing mass), as it can be appreciated in the central insert, where
a binning of 1 GeV is used.
If such higher resolution can be achieved, one can possibly disentangle
the $h\ar b\bar b$ peak from reaction (\ref{hh}) (lower-right insert) 
whereas the $H\ar b\bar b$ one from process (\ref{wh})
remains overwhelmed by the background (its position is located by the 
arrow in the upper-right insert).
At larger energies, phase space effects do not
suppress any longer the production of the heaviest neutral Higgs scalar.
In fact, at $\sqrt s=500$ GeV,
for $\MA=140$ GeV and \tba, the resonances of all three neutral Higgses
of the \mssm\ can be recognised (see Fig.~5b). In general, 
note that, although such plots might be somewhat superfluous if one
wants to optimise the cuts in order to enhance \eett\ and reduce the background
(in fact, 
by the time of the advent of an NLC the values of $M_{\rm{Higgs}}$
might already be known), in contrast
their knowledge will be essential if one would
like to attempt studies of the gauge structure of the \mssm, as 
mentioned in footnote 6. Like for the case of top signals, we again treat
the case of charged Higgs signals at $\sqrt s=350$ and 500 GeV separately.
\vskip0.5cm
\noindent\underline{{\boldmath${\sqrt s=500}$}~{\bf{GeV}}}
\vskip0.25cm
\noindent
We continue this Subsection by discussing the potential of a
$\sqrt s =500$ GeV NLC in detecting
charged Higgses produced in \eeHH\ scatterings, in which one of the scalars
decays into $hW^{\pm }\ar b\bar bW^{\pm }$ and/or 
$tb\ar b\bar b W^\pm$.
In Figs.~6a--b we display the differential 
distributions in the invariant mass of the system $b\bar b W^\pm$ (i.e., 
the hadronic four-jet subset, as $W^\pm\ar$ jj), 
at $\sqrt s=500$ GeV and for $\MA=140,180$
and 220 GeV (both values of $\tb$ are considered). For all pseudoscalar
Higgs masses we give the results as produced by the full set of Feynman
diagrams, whereas in the case $\MA=140$ GeV
we also plot the two resonant $H^\pm$ contributions separately (upper
two windows). Note that, on the one hand, we consider $A$ masses larger 
than 140 GeV (i.e., $M_{\Hpm}\OOrd 161$ GeV) to allow for large
values of the two Higgs BRs into $hW^{\pm }$ and
$tb$ (either or both), on the other hand, we restrict ourself
to the case $\MA\Ord 220$ GeV because of the maximum CM energy available.
It is interesting to notice how the spectra produced by the non-$H^+H^-$
diagrams steeply increase at $M_{b\bar b W^\pm}\approx170-180$ GeV (roughly,
the top rest mass: compare
the lower window in Fig.~6a against the two upper ones). This is somewhat
fortunate, as for $\MA=140$ GeV the charged Higgs peak lies right a few
GeV below that. Furthermore,  for this value of $\MA$ and for $\tba$, the
Higgs production and decay rates are similar in the two channels (compare
the continuous curves in the two upper frames of Fig.~6a). This could
allow one 
(after implementing an appropriate selection requirement: e.g., 
$M_{b\bar b}\approx M_h$) to separate the $hW$ decay from the $tb$ one.
In fact, for larger values of $M_{\Hpm}$ the Higgs
signals come in practise from top-bottom decays only (see Fig.~6b). For
$\tbb$ this is also true at $\MA\approx140$ GeV \cite{ioejames}, as it can be 
appreciated by comparing the dashed curves in the two upper sections
of Fig.~6a. 

Since the rates given in Figs.~6a--b do not include the BR into
$\tau\nu_\tau$ pairs of one of the $H^\pm$ bosons and since this becomes 
very small for $\MA\OOrd140$ GeV at small \tb's \cite{ioejames}, 
one has to face the following scenario (assuming
the nominal integrated luminosity $\int{\cal L}dt=10$ fb$^{-1}$). 
If \tb\ is large,
charged Higgs signals can be produced at large rate at an NLC with
$\sqrt s=500$ GeV and
certainly detected in the $H\ar tb\ar b\bar b W^{\pm}$ channel  
for $M_{\Hpm}\OOrd161$ GeV.  In contrast, 
for small $\tb$'s, this is true only for $\MA\approx140-150$ GeV, region
where the `narrow' contribution from the $hW$ channel compensates 
the small below threshold branching ratio into $tb$ pairs.
\vskip0.5cm
\noindent\underline{{\boldmath${\sqrt s=350}$}~{\bf{GeV}}}
\vskip0.25cm
\noindent
Concerning smaller values of CM energy, that is $\sqrt s=350$ GeV,
it must be said that phase space suppression largely counterbalances
the increase of the production cross section due to the $s$-channel
dynamics of \eeHH\ events. In the very end the spectrum of charged Higgs 
masses which can be covered by the two mentioned decay channels at that
energy is rather narrow, about 20 GeV, and only if \tba. In fact, 
for such a value, 
when $M_{H^\pm}$ is below 130 GeV or so, the total BR into the two Higgs 
channels is smaller than $0.1\%$. When instead $M_{H^\pm}\OOrd150$ GeV the
production cross section fall below detection level \cite{Higgsreview}. 
If \tbb\ such a window disappears completely. For these reason then, 
we decided not to focus our attention on the threshold energy stage of the
NLC, in the case of charged Higgs searches via \eeHH.

\subsection*{3.4 Mass dependence}

So far, we have concentrated in our analysis on the
case of selected values of \MA\ (or, equivalently, of \MHpm).
However, as the effects due to the finite widths and to the irreducible 
backgrounds are equally present over all the available charged Higgs mass 
range, in this closing Subsection we intend to generalise our results
by presenting various rates for processes (\ref{wh})--(\ref{hh})
as a function of \MHpm\ (again, at fixed \tba\ and 30.).
We do so, e.g., at the CM energy value of $\sqrt s=500$ GeV, as we have
noticed that a second stage NLC allows one to study 
contemporaneously both \eett\ and \eeHH\ events.

Fig.~7 shows the cross sections for the same three processes 
as defined in Tab.~I--II, for both reactions (\ref{wh})--(\ref{hh}),
but now at $\sqrt s=500$ GeV and over 
the continuous mass range 60 GeV $\Ord\MA\Ord$ 240 GeV 
(i.e., 100 GeV $\Ord\MHpm\Ord$ 252 GeV).
A comparison between the rates obtained in 
NWA (i.e., continuous lines) and those from the 
production and decay diagram only
(i.e., dashed lines) illustrates 
that finite width effects clearly becomes larger
as the difference $\Delta M=m_t-\MHpm-m_b$ gets smaller, the latter
being positive. If negative, the rates of the NWA are identically zero.
For $\MHpm$ well below $m_t$ the effects are negligible (also compare
to Figs.~2a--b at $\sqrt s=350$ GeV). Furthermore, as 
the top width does not vary drastically, from approximately 1.72(1.78) GeV at 
$\MA=60$ GeV and $\tba(30.)$ down to around 1.40 GeV at $\MA=150$ GeV,
where only the \sm\ decay channel $t\ar bW^\pm$ 
survives (see Fig.~2b of Ref.~\cite{io}),
width effects are (above the $t\ar bH^\pm$ threshold)
rather independent from $M_{H^\pm}$. 
The total rate of process (\ref{wh}) is, in comparison
to that of \eettbbwh\ events, much larger than the cross section of reaction
(\ref{hh}) is with respect to the corresponding \eettbbhh\ rates.
As already outlined in Section 3.1, this is due to the much richer
variety of non-$t\bar t$ events that
are active in the first case (especially the charged Higgs resonances:
notice, e.g., the onset of the top-bottom channel at large $\MHpm$'s
and $\tbb$). 

In Fig.~8 we have plotted the cross sections of the two possible 
Higgs mechanisms entering in \eebbwh\ events: i.e., via $H^\pm\ar tb$ and
$H^\pm\ar hW^\pm$ decays (actually, the second channel
is only considered at small $\tb$'s, because of the vanishing
BR at larger values \cite{ioejames}).
For both channels, the figure shows rates as 
obtained by multiplying the total \eeHH\ production rates times the 
branching ratios BR$(H^\pm\ar tb)$ and BR$(H^\pm\ar hW^\pm)$
(i.e., assuming on-shell production and neglecting the Higgs widths) and
those produced by the subamplitudes involving the two mentioned resonances
in our ME, along with the yield of all \eebbwh\ diagrams. 
Whereas the effects of the finite width of the $H^\pm$-bosons are generally
small in both Higgs channels (possibly apart from the $H^\pm\ar hW^\pm$
channel for $\MHpm\approx m_t$ at \tba), 
the irreducible background is overwhelming
the signals up to $\MHpm\approx150$ GeV (that is, as long as the $t\ar bH^\pm$
channel is well open), the rates of the
two remaining comparable up to charged Higgs
masses around 200 GeV. Above $\MHpm\approx150$ GeV, 
it is the $H^\pm\ar tb$ component of
the \eebbwh\ cross sections which accounts for the total rate, this being
modulated at large values of $\MHpm$ by a phase space suppression.
Since the largest part of the cross section of 
non-Higgs diagrams in \eebbwh\ events is due to $t\bar t$ events, the
implementation of a cut like $M_{jjj}\neq m_t$    
(on the three-jet invariant mass distributions that can be reconstructed
from the $b\bar b\rm{jj}\tau\nu_\tau$ signature) should allow one
to largely increase the Higgs signal-to-background ratio. However, notice that
in Section 3.3 no such cut was enforced, the Higgs peaks still 
being visible in many instances. This makes clear that
concrete chances of Higgs detection exist for charged scalars of the \mssm\
produced via \eeHH\ interactions and decaying via either $H^\pm\ar hW^\pm$
or $H^\pm\ar tb$ at $\sqrt s=500$ GeV $e^+e^-$ Linear Colliders. 

\section*{4. Summary and conclusions}

In this paper we have studied the  processes
$$
e^+e^-\ar b\bar b W^\pm H^\mp
\qquad\qquad\mbox{and}\qquad\qquad
e^+e^-\ar b\bar b H^+ H^-,
$$
by computing their complete matrix elements at tree-level and integrating
these over the corresponding phase spaces to produce total and differential 
cross sections relevant to phenomenological studies at the NLC.
We focused our attention on the CM energies of $\sqrt s=350 $ GeV (top-antitop
threshold, first stage NLC) and 500 GeV (final stage NLC). 

These processes are experimentally interesting since they both include
among the various contributions the production of top-antitop
pairs eventually decaying via the channels $t\ar bW^\pm$ and $t\ar bH^\pm$. 
Furthermore, in the case of \bbwh\ final states one can also study the 
charged Higgs production process \eeHH\ followed by the decay of one of the 
$H^\pm$'s via $hW$ and/or $tb$ pairs, with $h\ar b\bar b$ and $t\ar bW$.    

Numerical analyses have been carried out in order to quantify, first,
the influence of finite top width and irreducible
background effects on the integrated and differential rates as obtained from 
\eett\ events in NWA and, second,
to establish the detectability of the mentioned Higgs processes, as a function 
of the values assumed by the fundamental parameters of the \mssm.

For certain combinations of $\MA$ and $\tb$, our results indicate that, 
on the one hand, finite width and background effects must be included in the
simulations aiming to measure precisely the top parameters and, on the other
hand, charged Higgs signal in the mass range 140 GeV $\Ord M_{H^\pm}\Ord
\sqrt s/2$ can be promptly detected at nominal luminosity, especially if
$\sqrt s$ and \tb\ are large.

The numerical codes we have used in our analysis can be easily folded
with the experimental Monte Carlo simulation programs (including beam and 
detector effects) used by the NLC Working Groups, in order to quantify 
correctly irreducible background effects in top studies as well as 
to optimise the selection strategy of charged Higgs signals. In this respect, 
note that we have always assumed one 
of the two bosons in the final state to decay into $\tau\nu_\tau$ pairs, which
is clearly restrictive in the case of Higgs searches for
large values of $\MA$ and small $\tb$'s.
Indeed, dedicated experimental simulations might well assess that other 
$H^+H^-$ decay signatures are also relevant, so that we make our matrix 
elements available to the public for further studies.  

Finally, 
we also have pointed out the importance of events of the type discussed
here in order to study the gauge structure of the Higgs sector of the \mssm,
as they involve many of the tree-level Higgs vertices 
of the theory.

\section*{5. Acknowledgements}

SM is grateful to the UK PPARC and
KO to Trinity College and the Committee of Vice-Chancellors and
Principals of the Universities of the United Kingdom for
financial support.
This work is sponsored in part by the EC Programme
``Human Capital and Mobility'', Network ``Physics at High Energy
Colliders'', contract CHRX-CT93-0357 (DG 12 COMA).

\goodbreak

\vfill
\newpage

\subsection*{Table Captions}
\begin{description}

\item[{\bf Tab.~I}  ] Cross sections in femtobarns
for $e^+e^-\ar t\bar t$ (NWA, $\Gamma\ar0$), for
$e^+e^-\ar t\bar t\ar b\bar b W^\pm H^\mp$ (production and decay diagram
only, $\Gamma=\Gamma_t$) and for \eebbwh\ (all diagrams at tree-level,
$\Gamma=\Gamma_t$), 
at $\sqrt s=350$ GeV, for $m_t=174$ GeV, 
and $\MA=60(140)[220]$ GeV (i.e., $M_{H^\pm}\approx100(161)[234]$). 
The first(second) row reports rates for \tba(30.).
 
\item[{\bf Tab.~II}  ] Cross sections in femtobarns
for $e^+e^-\ar t\bar t$ (NWA, $\Gamma\ar0$), for
$e^+e^-\ar t\bar t\ar b\bar b H^+ H^-$ (production and decay diagram
only, $\Gamma=\Gamma_t$) and for \eebbhh\ (all diagrams at tree-level,
$\Gamma=\Gamma_t$), 
at $\sqrt s=350$ GeV, for $m_t=174$ GeV, 
and $\MA=60(140)[220]$ GeV (i.e., $M_{H^\pm}\approx100(161)[234]$). 
The first(second) row reports rates for \tba(30.).
 
\end{description}

\vfill
\newpage

\subsection*{Figure Captions}
\begin{description}

\item[{\bf Fig.~1} ] The \mssm\ Feynman topologies contributing at lowest order
to process (\ref{wh}) ({\bf a}) and (\ref{hh}) ({\bf b}). The label `h(h')' 
represents any of the neutral Higgses of the theory, $H, h$ and $A$, whereas
`H' refers to the charged scalar, $H^\pm$. `A', `Z' and `W' represent the
gauge bosons $\gamma, Z$ and $W^\pm$. Note that for process
(\ref{wh}) we plot the diagrams for the charge combination $W^-H^+$ only,
those corresponding to the case $W^+H^-$ being trivially deducible from
the one given here. The figure has been produced
with the help of the package MadGraph \cite{tim}. For reference,
in ({\bf c}) we present the branching ratios of $t$-quarks and $H^\pm$-bosons
as functions of the charged Higgs mass in the range 
$60~\mbox{GeV}\Ord M_{H^\pm}\Ord240$ GeV, for $\tba$ and 30.

\item[{\bf Fig.~2} ] Cross section in femtobarns for \eebbwh\ ({\bf a}) 
and \eebbhh\ ({\bf b}) events
around the top-antitop threshold, for three
different values of $\MA$. 
Continuous lines: \tba\ and 
production and decay diagram only. 
Dashed lines: \tba\ and
all diagrams. 
Dotted lines: \tbb\ and 
production and decay diagram only. 
Dash-dotted lines: \tbb\ and
all diagrams. In all cases $\Gamma=\Gamma_t$ has been used. 
Note the superposition of curves in some instances.

\item[{\bf Fig.~3} ] Differential distributions in the momentum of
the $W^\pm$ and $H^\pm$ bosons, in \eebbwh\  and 
\eebbhh\ events, for two values of $\MA$, 
with $\tan\beta=1.5$ ({\bf a}) and 30. ({\bf b}).
The two following cases are considered:
NWA (lower windows) and all diagrams (upper windows).
The CM energy is $\sqrt s=350$ GeV.
Continuous lines: $W^\pm$ momentum in \eebbwh\ events.
Dashed lines: $H^\pm$ momentum in \eebbwh\ events.
Dotted lines: $H^\pm$ momentum in \eebbhh\ events.

\item[{\bf Fig.~4} ] Differential distributions in the invariant
mass of the {`right'} and {`wrong'} three-jet combinations (see the text),
in \eebbwh\ and \eebbhh\ events, for $\MA=60$ (upper window)
and 100 (lower window) GeV, 
with $\tan\beta=1.5$ ({\bf a}) and $30.$ ({\bf b}). All diagrams have been here
considered. The CM energy is 
$\sqrt s=500$ GeV. 
Continuous lines: $M_{bW^\pm }$ mass in \eebbwh\ events.
Dashed lines: $M_{bH^\pm }$ mass in \eebbwh\ events.
Dotted lines: $M_{bH^\pm }$ mass in \eebbhh\ events.

\item[{\bf Fig.~5} ] Differential distributions in invariant mass
of the $b\bar b$ pair. ({\bf a}) In \eebbwh\ and \eebbhh\ events, 
for $\tan\beta=1.5$ GeV,
$\MA=60$ GeV and $\sqrt s=350$ GeV.
Continuous lines (main window): all \eebbwh\ diagrams.
Dashed lines (main window): all \eebbhh\ diagrams.
The two small windows on the right 
show the regions around the $H$ (in \eebbwh, upper one) and $h$ 
(in \eebbhh, lower one) resonances 
enlarged and plotted with high resolution, together 
with the $t\bar t$ contribution ($\Gamma=\Gamma_t$, shaded).
The small central window does the same for the $h$ and $A$ `resolved'
resonances in the case of \eebbwh\ events.
({\bf b}) In \eebbwh\ events, for $\tan\beta=1.5$ GeV,
$\MA=140$ GeV and $\sqrt s=500$ GeV.
Dashed line: all diagrams. Continuous line: 
$t\bar t$ contribution ($\Gamma=\Gamma_t$, shaded).

\item[{\bf Fig.~6} ] Differential distributions in invariant mass
of the $b\bar b W^\pm$ system, in \eebbwh\ events, 
for $\tan\beta=1.5$ and 30., $\MA=140$ ({\bf a}) and $180$, 220 GeV 
({\bf b}), at $\sqrt s=500$ GeV.
In the case $\MA=140$ GeV the resonant $H^\pm\ar hW^\pm$ (upper plot)
and $H^\pm\ar tb$ (middle plot) contributions are shown separately.
Continuous lines: \tba.
Dashed lines: \tbb\

\item[{\bf Fig.~7} ] Cross section in femtobarns for \eebbwh\ (upper three
curves) and \eebbhh\ (lower three curves) events at $\sqrt s=500$ GeV as
a function of $M_{H^\pm}$, for \tba\ (upper plot) and \tbb\ (lower plot).
Continuous lines: \eett\ (NWA, $\Gamma\ar0$).
Dashed lines: \eettbbwh\ and \eettbbhh\ (production and decay diagrams only, 
              $\Gamma=\Gamma_t$).
Dotted lines: \eebbwh\ and \eebbhh\ (all diagrams at tree-level, 
              $\Gamma=\Gamma_t$).

\item[{\bf Fig.~8} ] Cross section in femtobarns for \eebbwh\ 
events at $\sqrt s=500$ GeV as
a function of $M_{H^\pm}$, for \tba\ (upper plot) and \tbb\ (lower plot).
Continuous lines: $\sigma(e^+e^-\ar H^+H^-)\times
                   \mbox{BR}(H^\pm\ar t b\ar b\bar b W^\pm)$.
Dashed lines: \eeHH\ $\ar t b H^\mp\ar$ \bbwh.
Dotted lines: $\sigma(e^+e^-\ar H^+H^-)\times
               \mbox{BR}(H^\pm\ar h W^{\pm }\ar b\bar b W^\pm)$.
Double-dash lines: \eeHH\ $\ar h W^{\pm }H^\mp\ar$ \bbwh. 
Dash-dotted lines: \eebbwh\ (all diagrams at tree-level, $\Gamma=\Gamma_t$).

\end{description}

\vfill
\clearpage

\begin{table}%[p]%[htbp]
\begin{center}
\begin{tabular}{|c|c|c|c|}
\hline
\rule[0cm]{0cm}{0cm}
$m_t$ (GeV)                           &\omit  
$~$                                   &\omit  
$\sigma(e^+e^-\ar X)$ (fb)            &
$~$                                   \\ \hline  \hline
\rule[0cm]{0cm}{0cm}
$~$                                  &  
$t\bar t$                            &  
$t\bar t \rightarrow \bar b bW^\pm H^\mp$ & 
$\bar b bW^\pm H^\mp$                 \\ \hline\hline
\rule[0cm]{0cm}{0cm}
$174$ &  
$34.64(2.21)[1.32\times10^{-18}]$  &  
$30.55(1.73)[2.57\times10^{-8}]$   &  
$40.50(10.32)[9.48\times10^{-8}]$  \\
                             &  
$20.04(1.30)[6.37\times10^{-19}]$  &  
$17.57(1.01)[1.24\times10^{-8}]$   &  
$18.89(1.05)[9.33\times10^{-8}]$   \\ \hline\hline
\multicolumn{4}{|c|}
{\rule[0cm]{0cm}{0cm}
$\sqrt s=350$ GeV}
 \\ \hline 
\multicolumn{4}{c}
{\rule{0cm}{1cm}
{\Large Tab. I}}  \\
\multicolumn{4}{c}
{\rule{0cm}{0cm}}
\end{tabular}
\end{center}
\end{table}

\vskip2.0cm

\begin{table}%[p]%[htbp]
\begin{center}
\begin{tabular}{|c|c|c|c|}
\hline
\rule[0cm]{0cm}{0cm}
$m_t$ (GeV)                           &\omit  
$~$                                   &\omit  
$\sigma(e^+e^-\ar X)$ (fb)            &
$~$                                   \\ \hline  \hline
\rule[0cm]{0cm}{0cm}
$~$                                  &  
$t\bar t$                            &  
$t\bar t \rightarrow \bar b bH^+H^-$ & 
$\bar b bH^+H^-$                     \\ \hline\hline
\rule[0cm]{0cm}{0cm}
$174$ &  
$7.87(2.17\times10^{-2})[0.00]$  &  
$6.98(1.46\times10^{-2})[0.00]$  &  
$7.10(1.46\times10^{-2})[0.00]$  \\
                             &  
$2.84(7.50\times10^{-3})[0.00]$  &  
$2.49(5.04\times10^{-3})[0.00]$  &  
$2.52(5.05\times10^{-3})[0.00]$  \\ \hline\hline
\multicolumn{4}{|c|}
{\rule[0cm]{0cm}{0cm}
$\sqrt s=350$ GeV}
 \\ \hline 
\multicolumn{4}{c}
{\rule{0cm}{1cm}
{\Large Tab. II}}  \\
\multicolumn{4}{c}
{\rule{0cm}{0cm}}
\end{tabular}
\end{center}
\end{table}

\vfill
\newpage

\begin{figure}[p]
~\epsfig{file=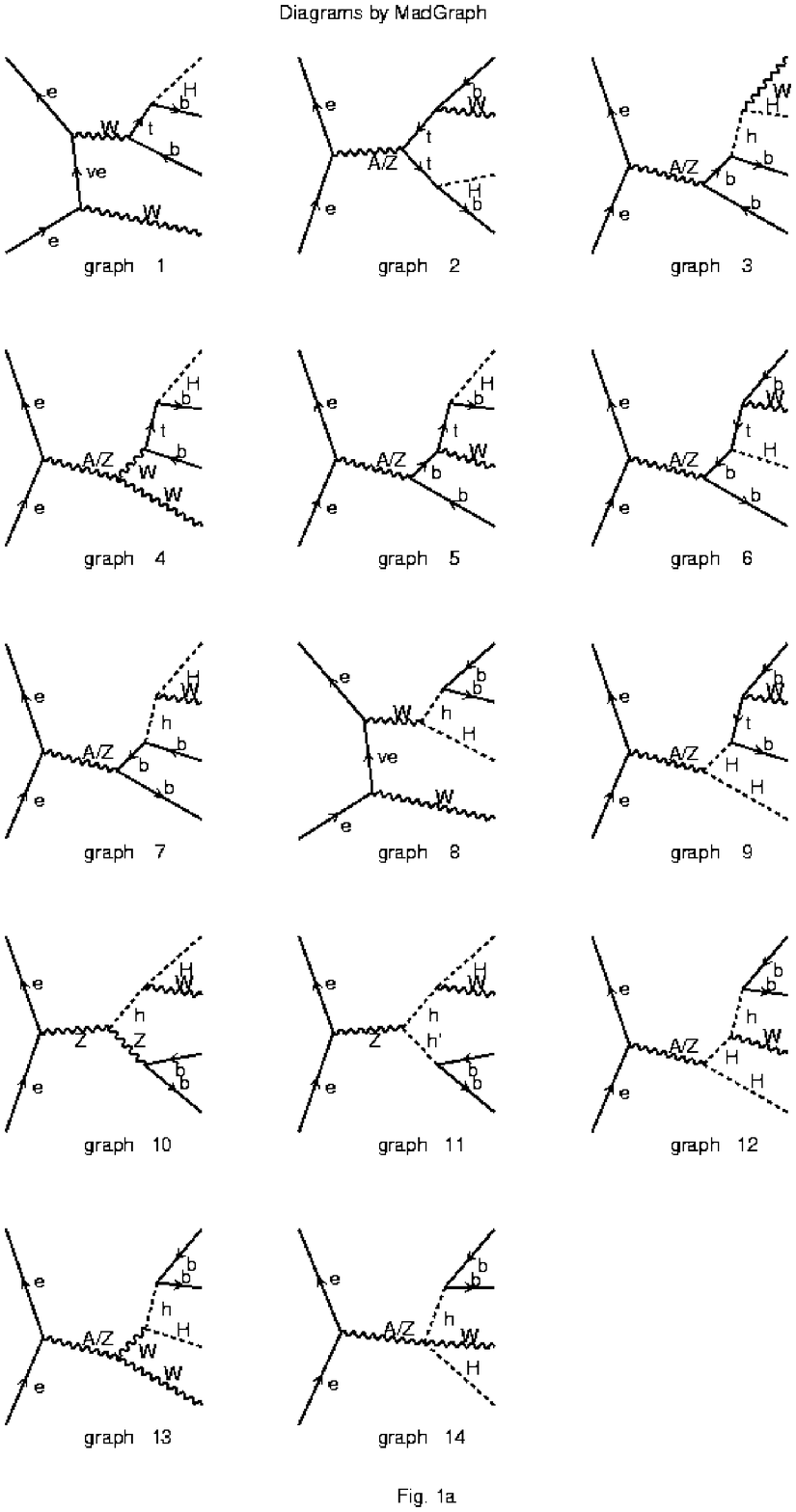,height=22cm}
\end{figure}
\stepcounter{figure}
\vfill
\clearpage

\begin{figure}[p]
~\epsfig{file=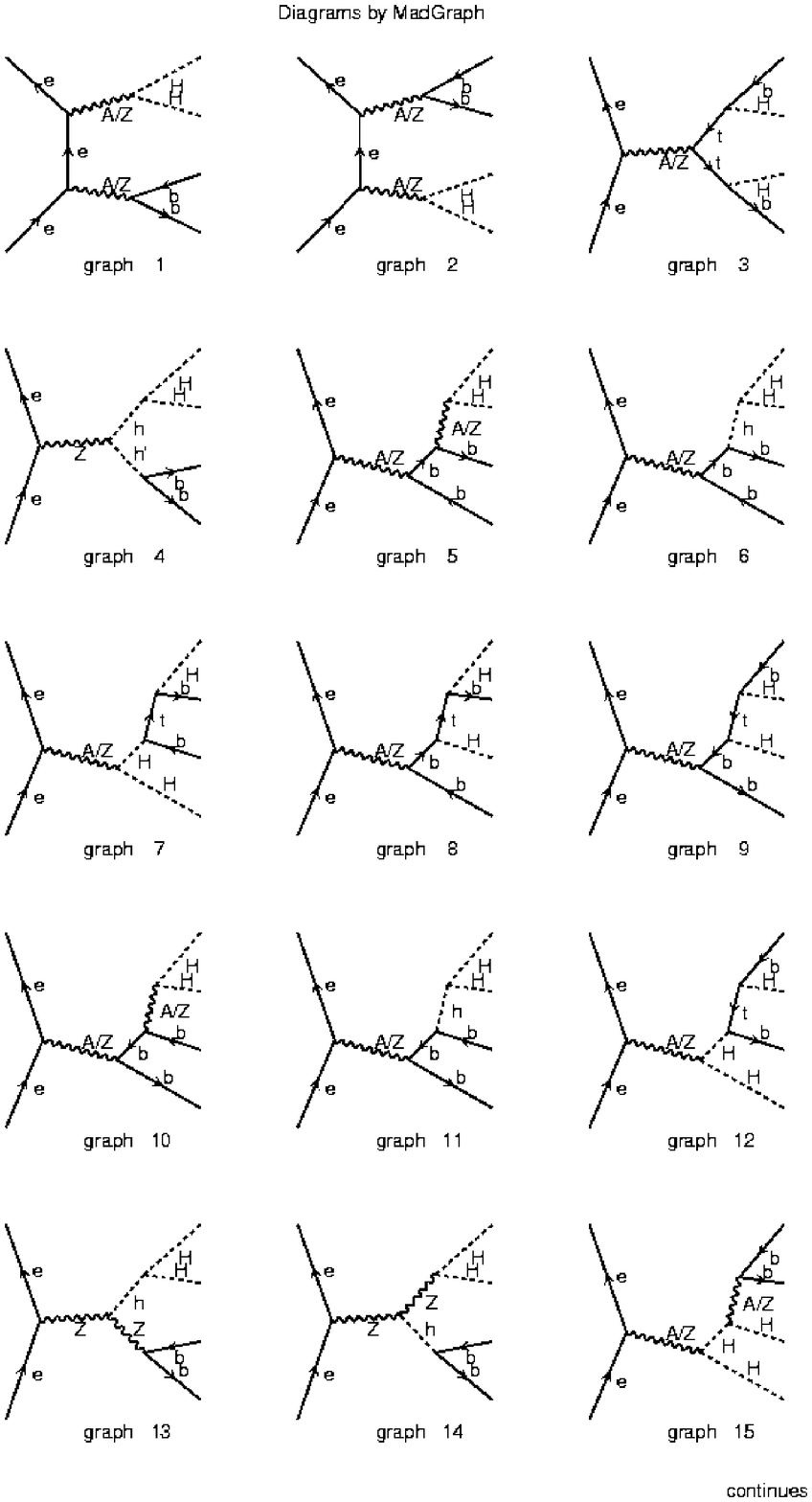,height=22cm}
\end{figure}
\stepcounter{figure}
\vfill
\clearpage

\begin{figure}[p]
~\epsfig{file=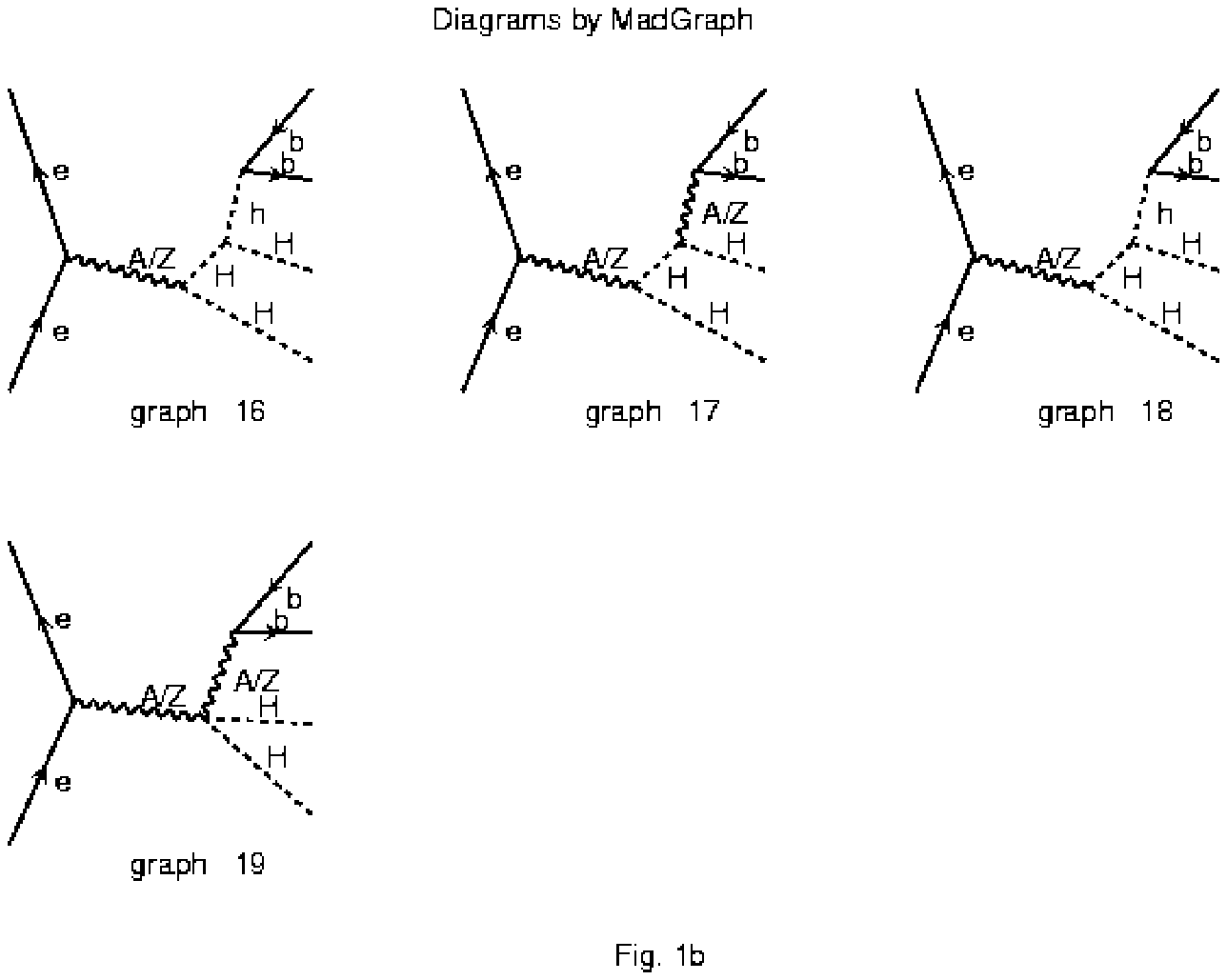,height=22cm}
\end{figure}
\stepcounter{figure}
\vfill
\clearpage

\begin{figure}[p]
~\epsfig{file=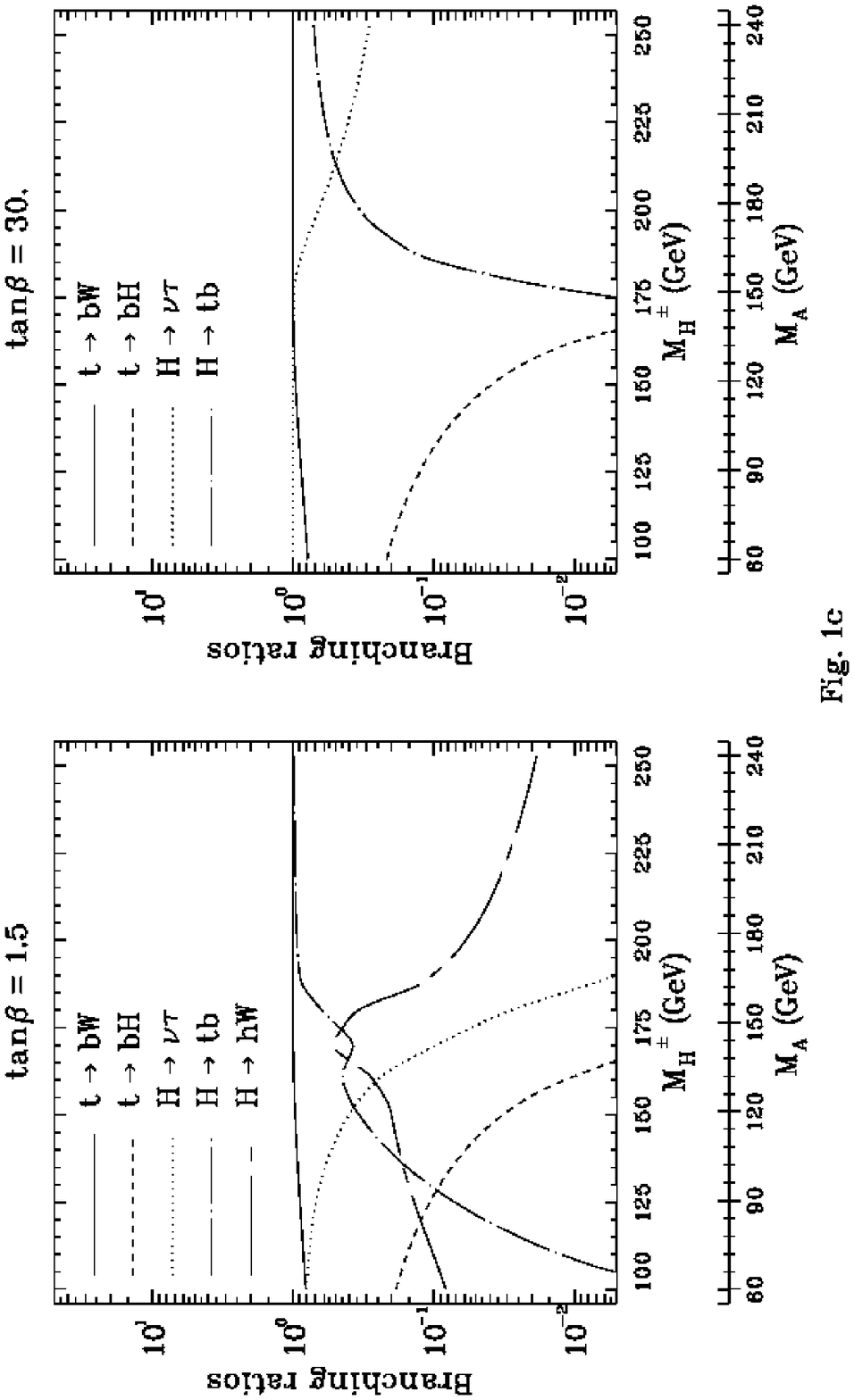,height=22cm}
\end{figure}
\stepcounter{figure}
\vfill
\clearpage

\begin{figure}[p]
~\epsfig{file=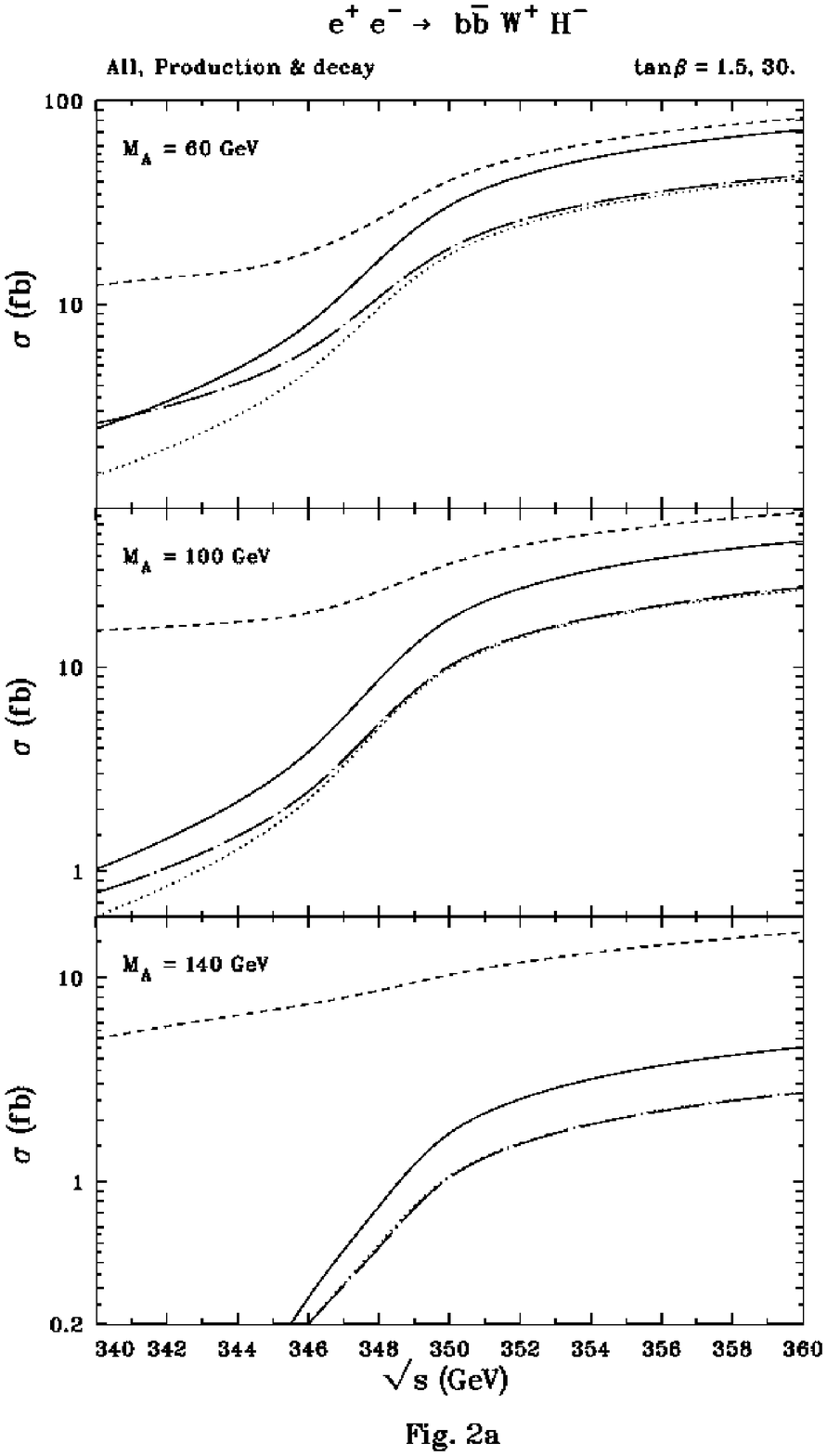,height=22cm}
\end{figure}
\stepcounter{figure}
\vfill
\clearpage

\begin{figure}[p]
~\epsfig{file=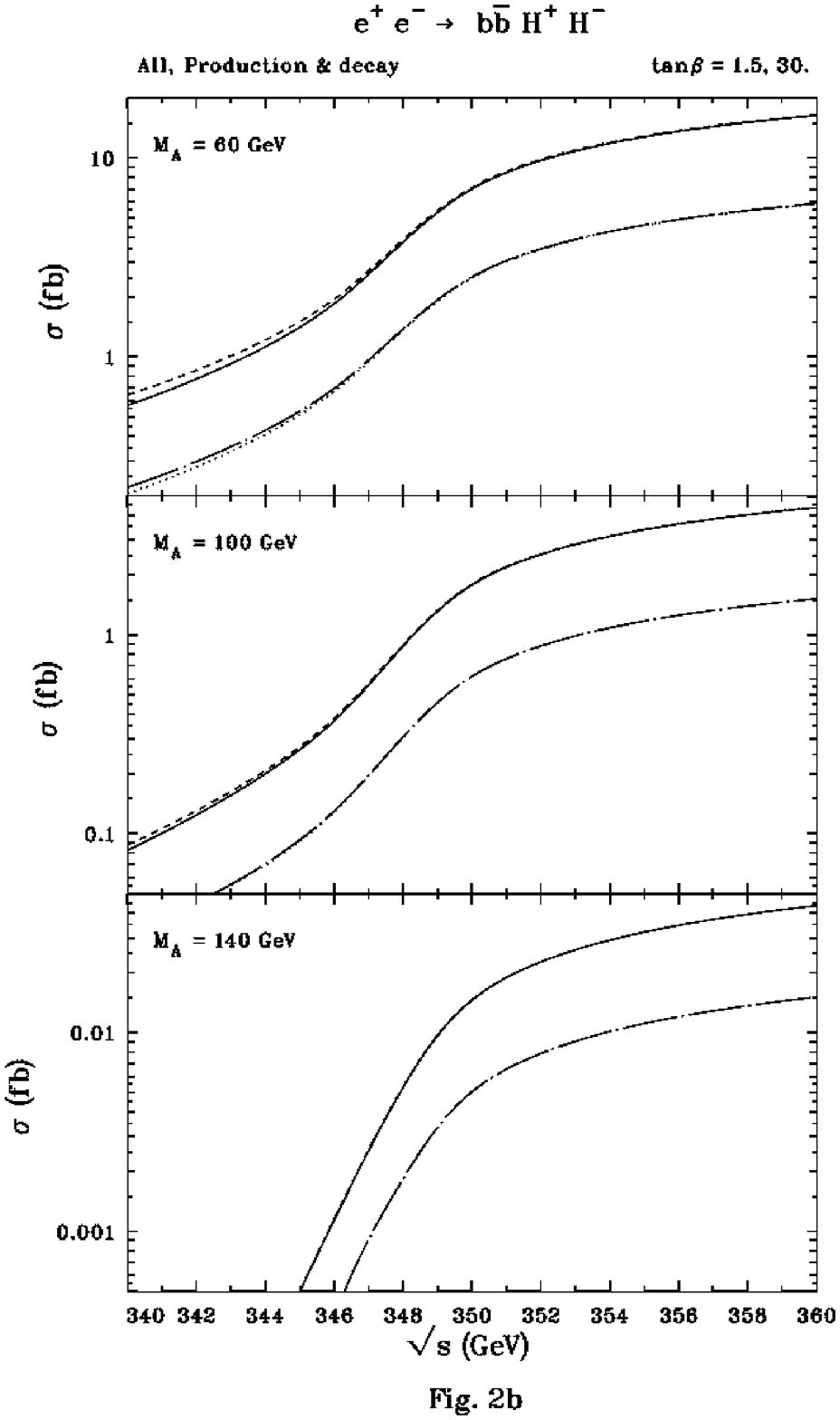,height=22cm}
\end{figure}
\stepcounter{figure}
\vfill
\clearpage

\begin{figure}[p]
~\epsfig{file=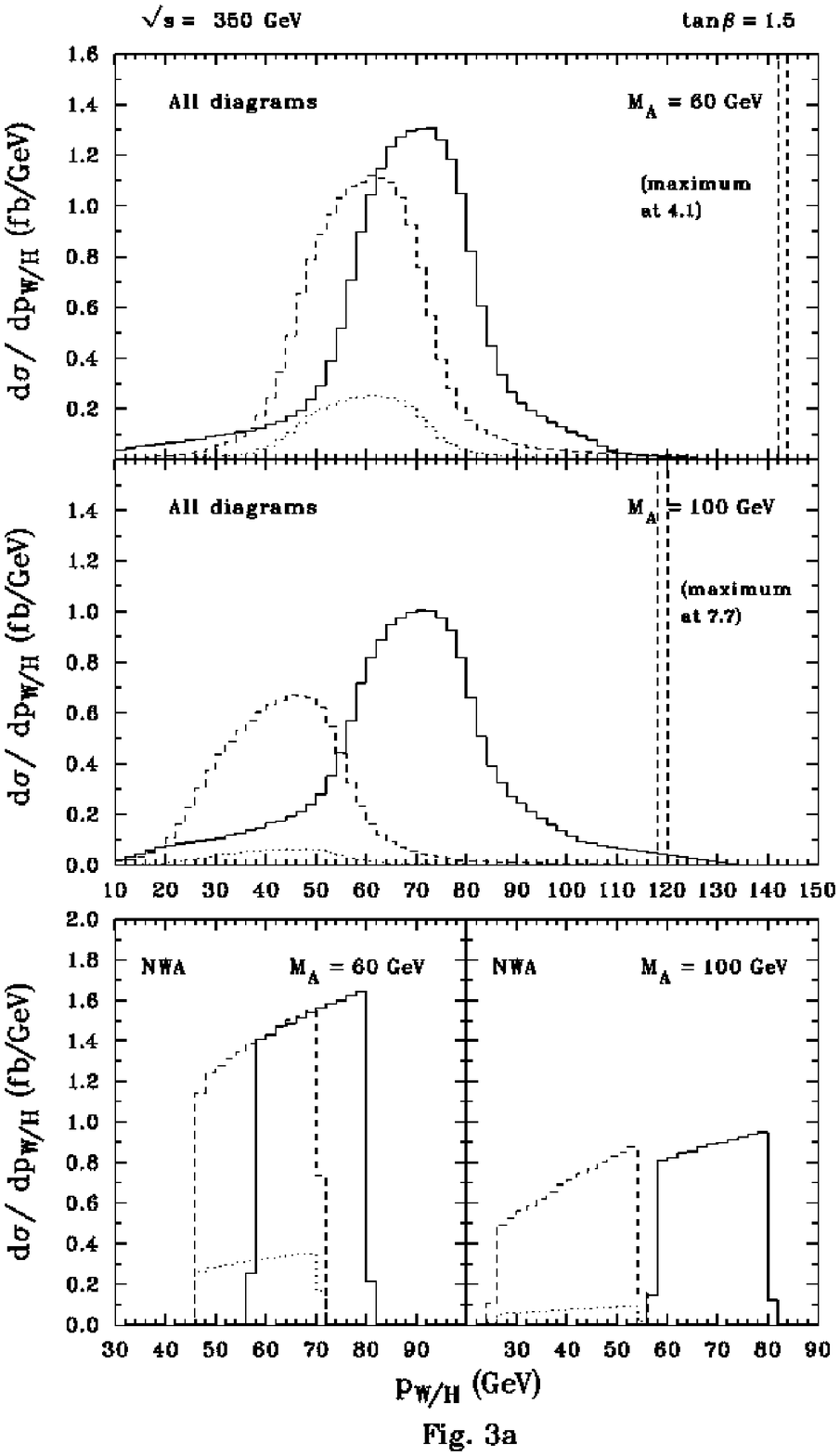,height=22cm}
\end{figure}
\stepcounter{figure}
\vfill
\clearpage

\begin{figure}[p]
~\epsfig{file=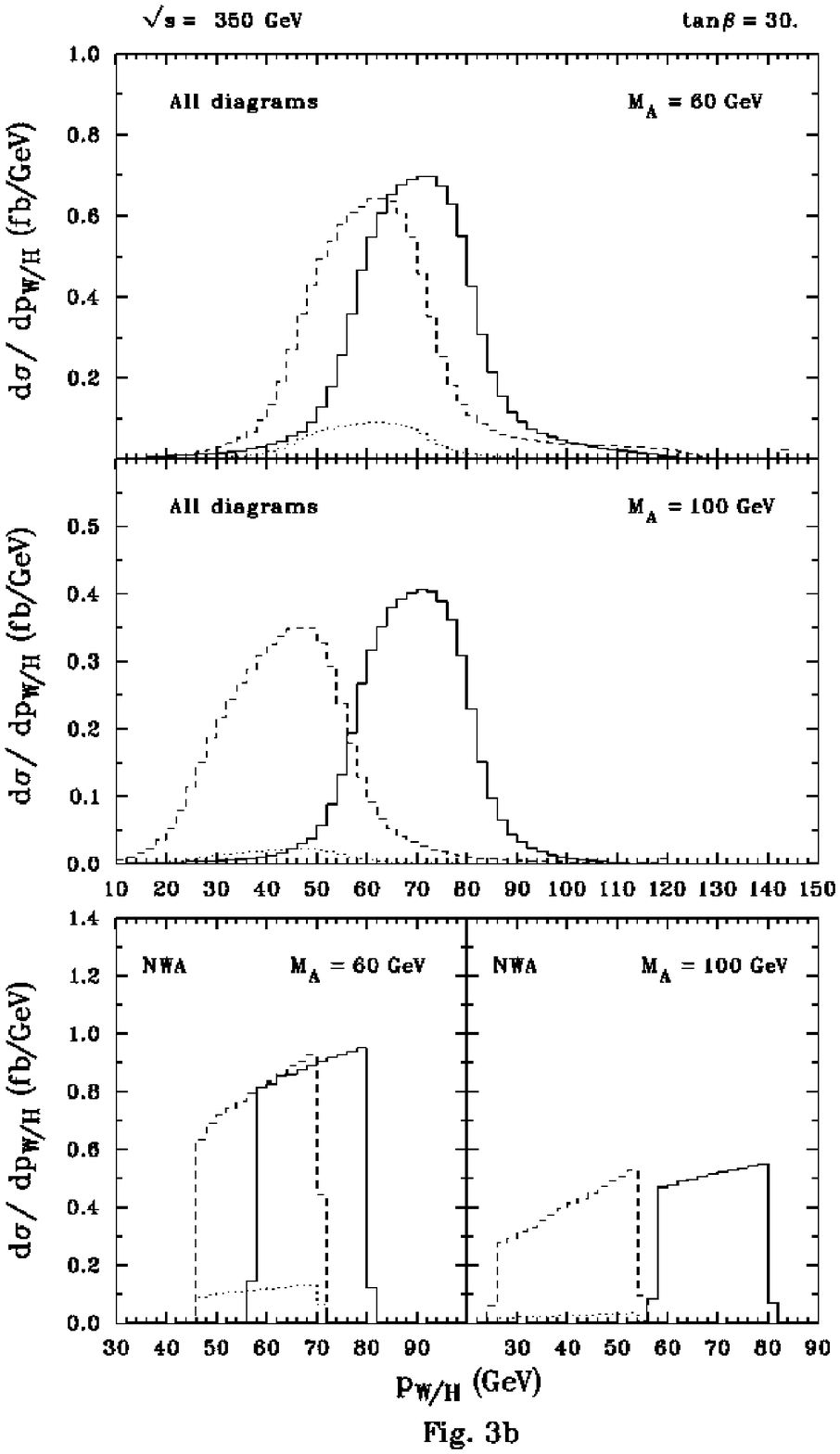,height=22cm}
\end{figure}
\stepcounter{figure}
\vfill
\clearpage

\begin{figure}[p]
~\epsfig{file=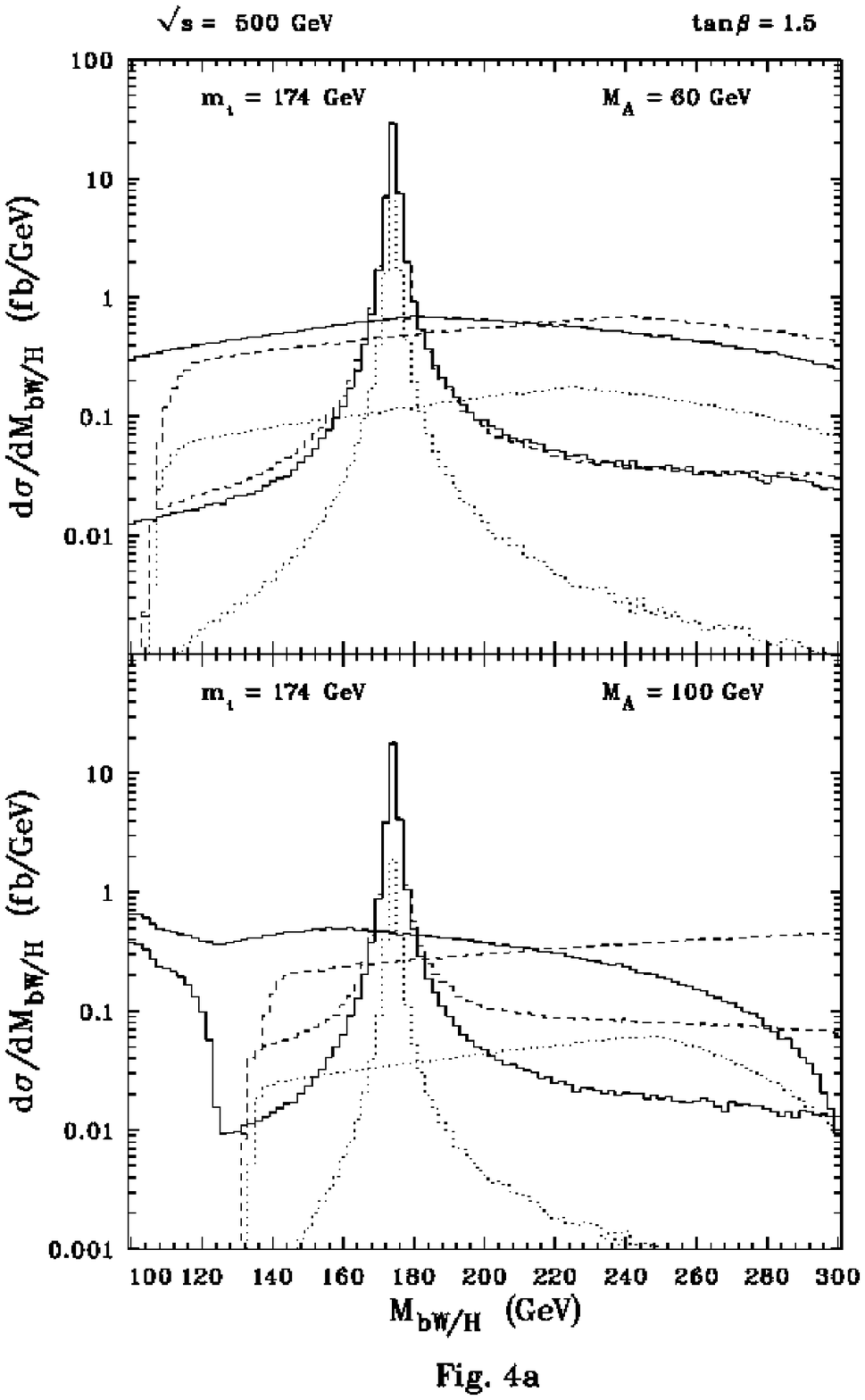,height=22cm}
\end{figure}
\stepcounter{figure}
\vfill
\clearpage

\begin{figure}[p]
~\epsfig{file=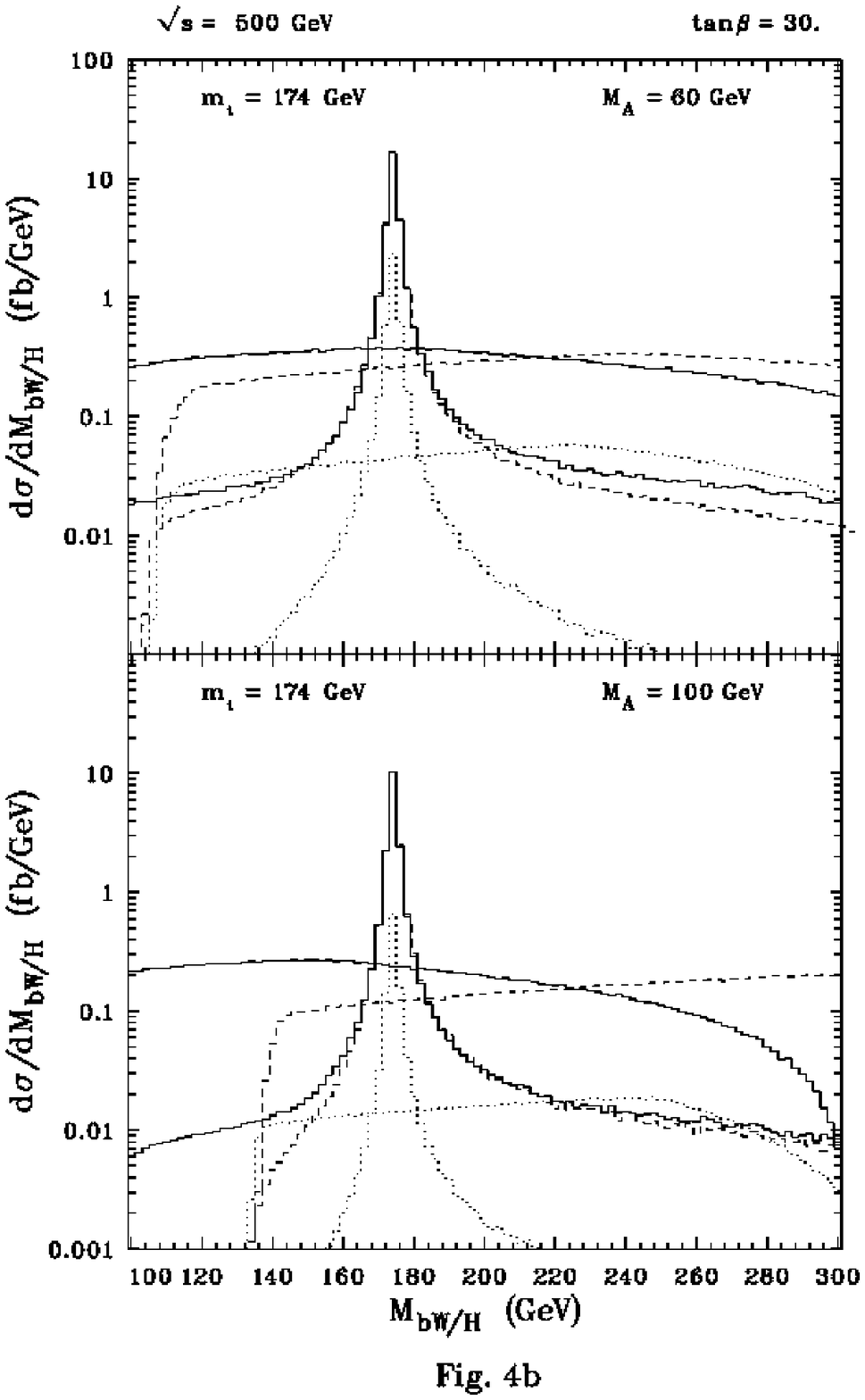,height=22cm}
\end{figure}
\stepcounter{figure}
\vfill
\clearpage

\begin{figure}[p]
~\epsfig{file=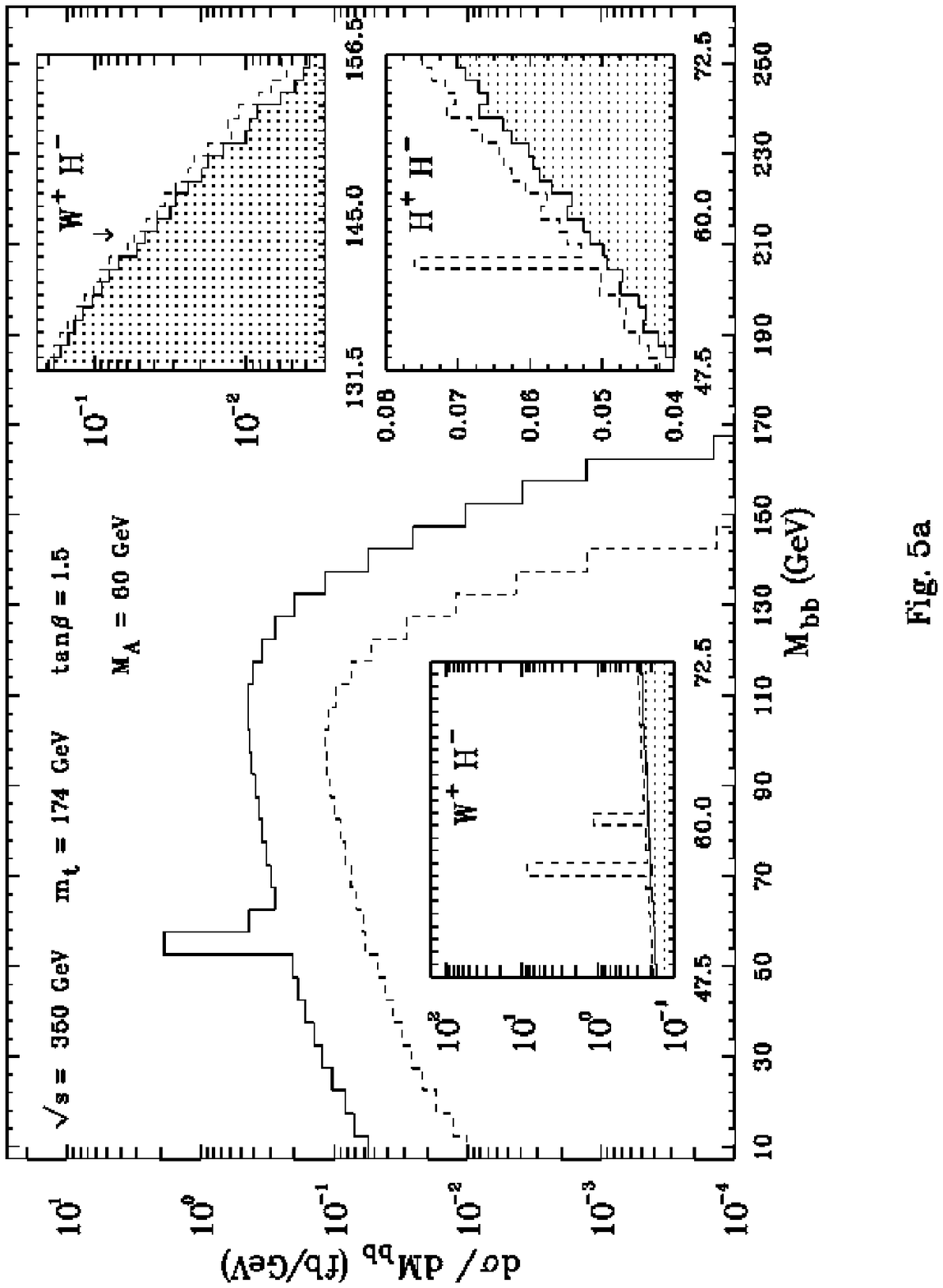,height=22cm}
\end{figure}
\stepcounter{figure}
\vfill
\clearpage

\begin{figure}[p]
~\epsfig{file=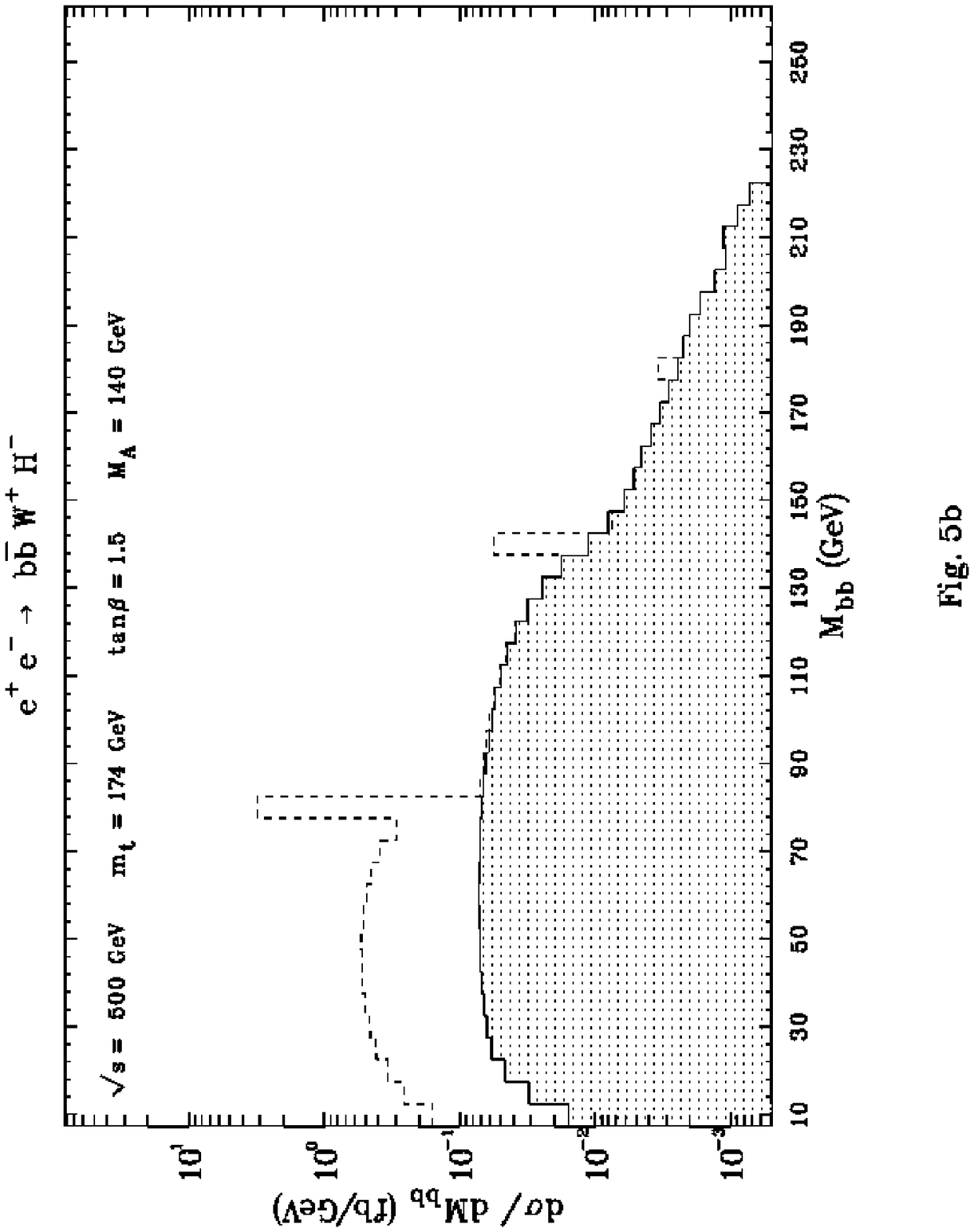,height=22cm}
\end{figure}
\stepcounter{figure}
\vfill
\clearpage

\begin{figure}[p]
~\epsfig{file=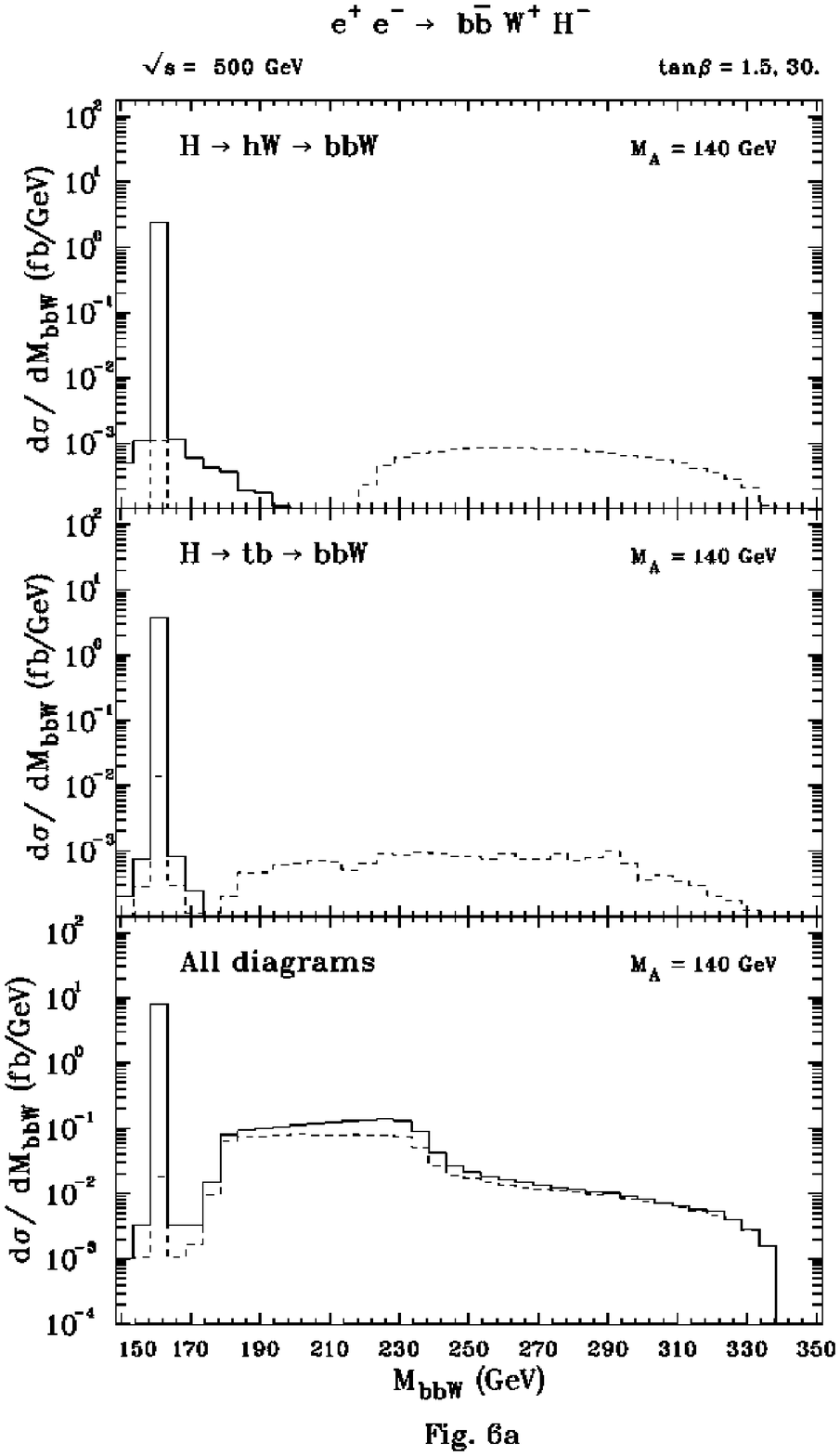,height=22cm}
\end{figure}
\stepcounter{figure}
\vfill
\clearpage

\begin{figure}[p]
~\epsfig{file=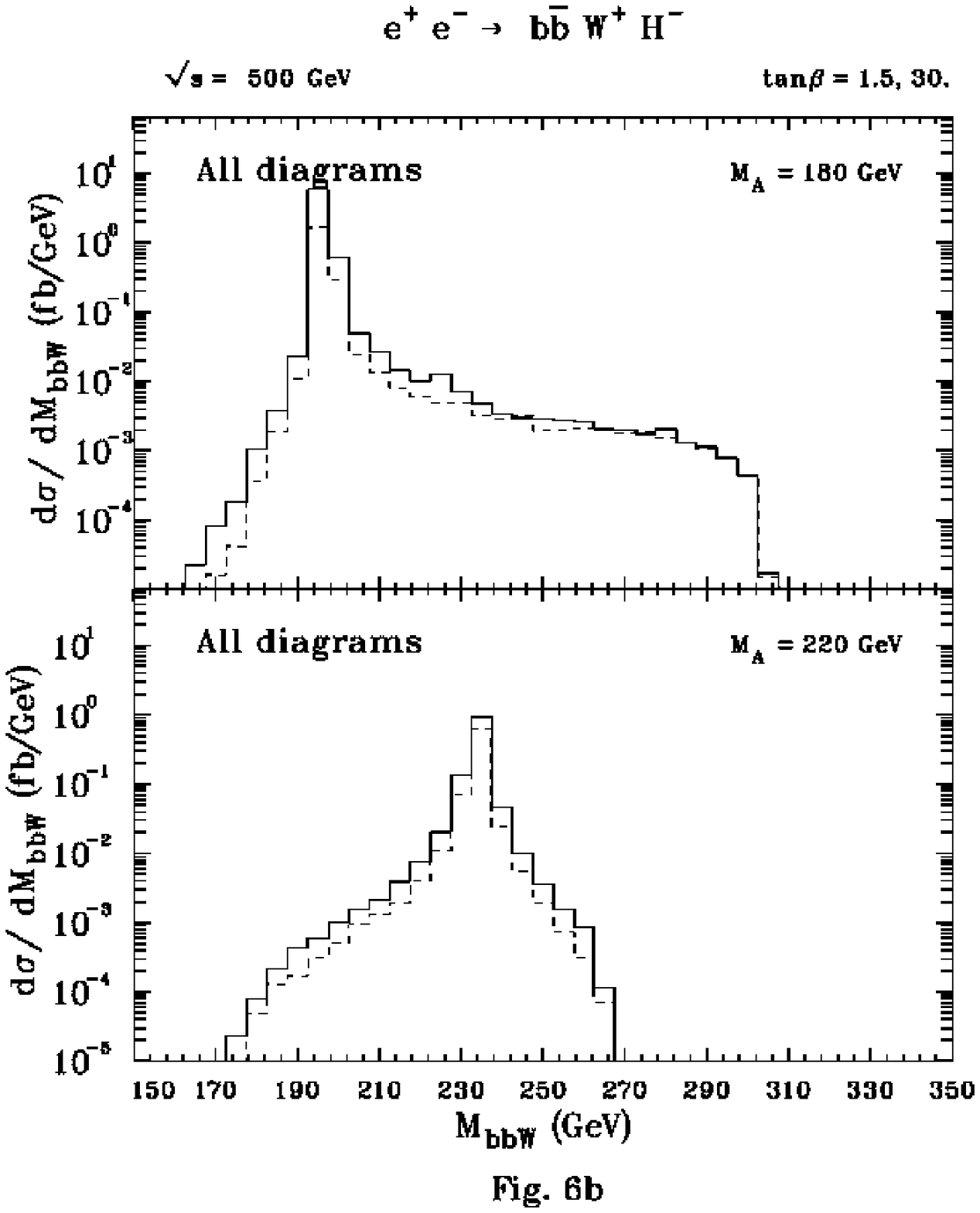,height=22cm}
\end{figure}
\stepcounter{figure}
\vfill

\begin{figure}[p]
~\epsfig{file=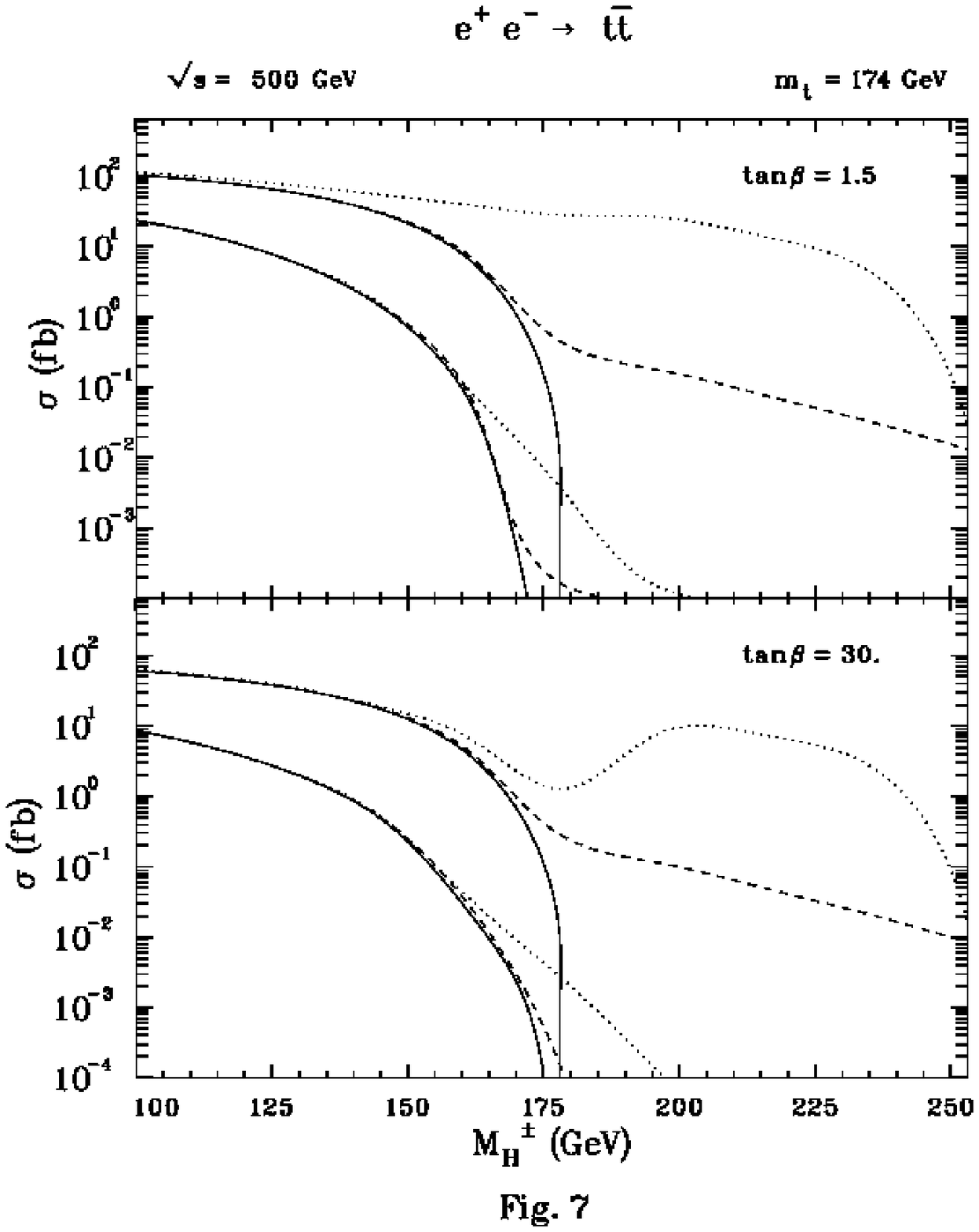,height=22cm}
\end{figure}
\stepcounter{figure}
\vfill

\begin{figure}[p]
~\epsfig{file=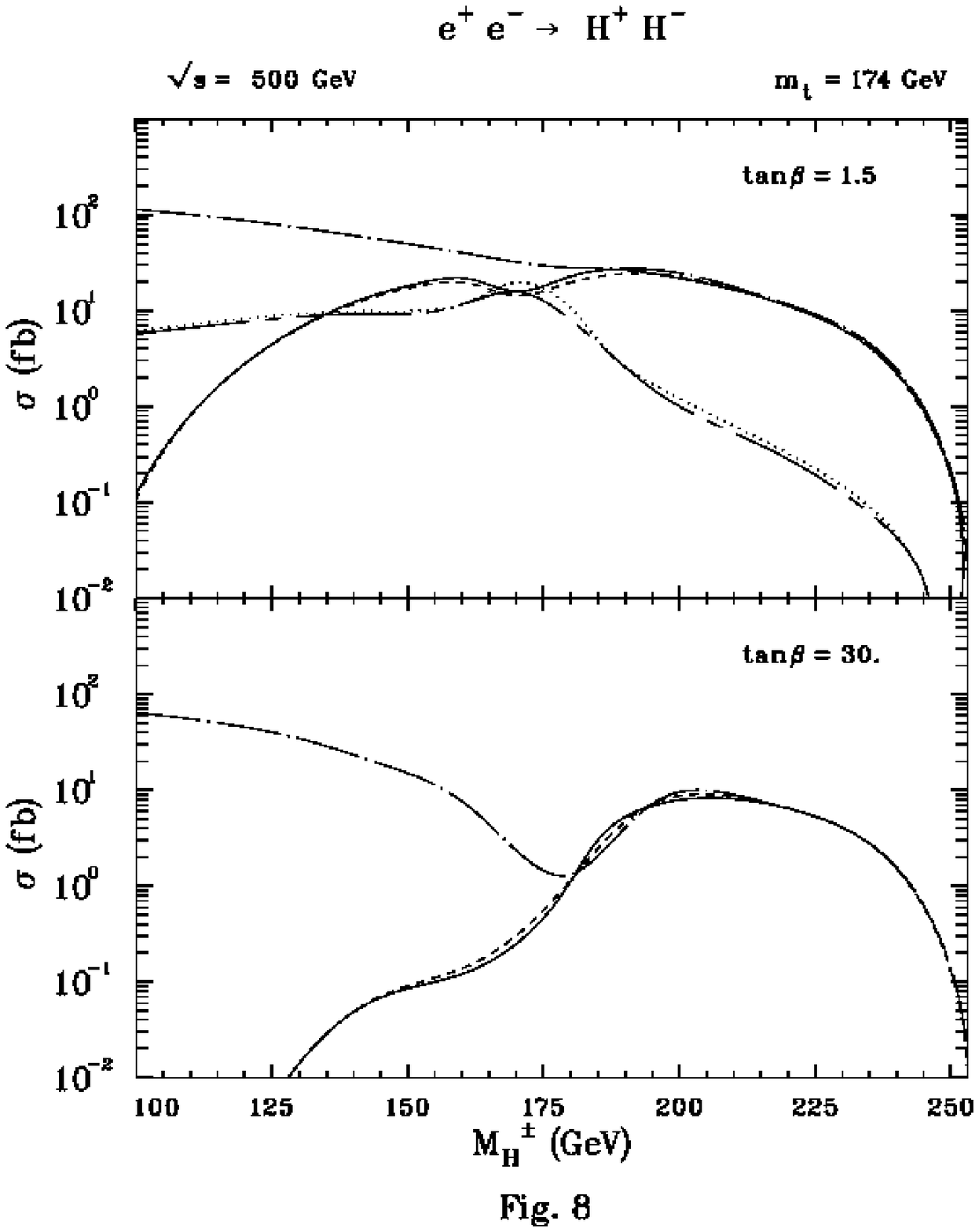,height=22cm}
\end{figure}
\stepcounter{figure}
\vfill

\end{document}